\documentclass[journal,twoside]{IEEEtran}
\usepackage{cite}
\usepackage{url}
\usepackage{amsmath,amssymb,amsfonts}
\usepackage{algorithmicx}
\usepackage{graphicx}
\usepackage{textcomp}
\usepackage{algorithm}
\usepackage{algpseudocode}

\DeclareMathOperator*{\argmin}{arg\,min}

\def\BibTeX{{\rm B\kern-.05em{\sc i\kern-.025em b}\kern-.08em
    T\kern-.1667em\lower.7ex\hbox{E}\kern-.125emX}}
\begin{document}
\title{Do CNNs solve the CT inverse problem?}
\author{Emil Y. Sidky, \IEEEmembership{Member, IEEE}, Iris Lorente, Jovan G. Brankov, \IEEEmembership{Senior Member, IEEE},
and Xiaochuan Pan, \IEEEmembership{Fellow, IEEE}
\thanks{This work is supported in part by NIH
Grant Nos. R01-EB026282, R01-EB023968, and R01-EB023969.
The contents of this article are solely the responsibility of
the authors and do not necessarily represent the official
views of the National Institutes of Health.}
\thanks{E. Y. Sidky  and X. Pan are with the Department of Radiology at The University of
Chicago, Chicago, IL, 60637
(e-mails: sidky@uchicago.edu, xpan@uchicago.edu).}
\thanks{I. Lorente and J. G. Brankov are with the Department of Electrical and Computer Engineering
at the Illinois Institute of Technology, Chicago, IL, 60616
(e-mails: irislorente@gmail.com, brankov@iit.edu).}}

\maketitle

\begin{abstract} 
Objective: 
This work examines the claim made in the literature that the inverse problem
associated with image reconstruction in sparse-view computed tomography (CT)
can be solved with a convolutional neural network (CNN). \\
Methods: 
Training and testing image/data pairs are generated in a dedicated breast CT
simulation for sparse-view sampling, using two different object models. The trained CNN is
tested to see if images can be accurately recovered from their corresponding
sparse-view data.
For reference, the same sparse-view
CT data is reconstructed by the use of constrained total-variation (TV) minimization (TVmin),
which exploits sparsity in the gradient magnitude image (GMI). \\
Results: 
Using sparse-view data from images either in the training or testing set,
there is a significant discrepancy between the image obtained with the CNN
and the image that generated the data. For the same simulated scanning conditions,
TVmin is able to accurately reconstruct the test image.\\
Conclusion: 
We find that the sparse-view CT inverse problem cannot be solved for the particular published
CNN-based methodology that we chose and the particular object model that we tested.
Furthermore, this negative result
is obtained for conditions where TVmin is able to recover the test images.\\
Significance: 
The inability of the CNN to solve the inverse problem associated with sparse-view
CT, for the specific conditions of the presented
simulation, draws into question similar unsupported claims being
made for the use of CNNs to solve inverse problems in medical imaging.
\end{abstract}


\section{Introduction}
\label{sec:introduction}
\IEEEPARstart{M}{uch} recent literature on computed tomography (CT) image
reconstruction has focused on data-driven approaches to ``solve'' the associated inverse
problem.  In particular, deep-learning with convolutional neural networks (CNN) is being
developed for image reconstruction from sparse-view projection data \cite{jin2017deep,zhu2018image,han2018framing}.
As a procedure for finding an image from sparsely-sampled projection data, there is no
fundamental logical problem in the use of CNNs. The claim, however, that CNNs {\it solve}
the associated inverse problem remains unsubstantiated in the literature.

In the sparse-view CT literature using CNNs, there is not clear evidence
that an associated inverse problem is being solved. There does not seem
to be a framework that makes it clear how to determine CNN parameters, network design,
and training sets to obtain an accurate inverse problem solution.
In fact, in the majority of sparse-view CT CNN literature there is not even a clear
statement of what is the inverse problem being solved.
The lack of rigorous study of CNNs as an inverse problem solver for CT has motivated us to perform
our own investigation.

We focus specifically on the problem of sparse-view CT image reconstruction,
because many of the published works on CNN-based image reconstruction address this
problem.  The only known framework that has been shown to solve
sparse-view CT inverse problems is sparsity-exploiting image reconstruction.
This type of image reconstruction has typically been formulated as a 
large-scale non-smooth convex optimization,
where a substantial theoretical framework has been built-up over the past few years.

We design a simulation based on dedicated breast CT \cite{boone2001dedicated}, where minimizing X-ray
dose is of paramount importance and the question of reducing the necessary number of projections provides
an avenue for achieving this dose reduction.
Exploiting gradient-sparsity has proven particularly effective,
and we employ constrained total-variation (TV) minimization to show inversion of the sparse-view
CT inverse problem and to provide a reference for the CNN-based investigation.
Using a stochastic 2D breast model phantom, we generate ideal data for training the CNN to solve the sparse-view
CT inverse problem.

In this work, we define the sparse-view CT inverse problem, explain the evidence that TVmin solves this inverse
problem, and demonstrate that we are unable to solve this inverse problem using a published CNN technique
that claims to do exactly that.
A brief synopsis of inverse problems is given in Sec. \ref{sec:ip}. 
The measurement model and model inverse for sparsity-exploiting image reconstruction is presented
in Sec. \ref{sec:sparseview}.  Sparse-view image reconstruction with CNN-based deep-learning
is outlined in Sec. \ref{sec:CNN} from an inverse problem perspective. The object model for the breast
CT simulation is specified in Sec. \ref{sec:phantom}. Results for the sparse-view image reconstruction 
using TVmin and the CNN are shown in Sec. \ref{sec:results}. Finally,
we conclude in Sec. \ref{sec:conclusion}.

\section{Solving an inverse problem}
\label{sec:ip}

What is meant by ``solving an inverse problem'' requires explanation, and we follow the
notation of \cite{bal2012introduction}.
Inverse problems start with specifying a measurement/data model
\begin{equation*}
y = \mathcal{M}(x),
\end{equation*}
where $\mathcal{M}$ is an operator that yields data $y$ given model parameters $x$.
An important part of the measurement model also involves specification of any constraints on $x$.

The investigation of the measurement model begins
with addressing measurement model injectivity; namely, that the measurement
model is one-to-one,
\begin{equation}
\label{injectivity}
\mathcal{M}(x_1) =\mathcal{M}(x_2) \Rightarrow x_1 = x_2 .
\end{equation}
In words, if two measurements in the range of $\mathcal{M}$ are the same then
they correspond to the same parameter set. As an example, if $\mathcal{M}$ is a
linear transform then injectivity is obtained if there are no non-trivial null vectors
on the right of $\mathcal{M}$.

Once injectivity is established,
an inverse to the measurement model $\mathcal{M}^{-1}$ can be constructed.
It is then logical
to address stability of the inverse of $\mathcal{M}$; namely, if the difference
between two data vectors $y_1$ and $y_2$ is small then the difference between
their corresponding parameter estimates $\mathcal{M}^{-1}(y_1)$ and
 $\mathcal{M}^{-1}(y_2)$ is also small
\begin{equation}
\label{stability}
\|\mathcal{M}^{-1}(y_1) - \mathcal{M}^{-1}(y_2)\| \leq \omega ( \|y_1 - y_2\|),
\end{equation}
where $\omega (\cdot)$ is a monotonic function, mapping positive reals to positive reals and $\omega (0)=0$.
The measurement model inverse $\mathcal{M}^{-1}$
is stable if $\omega (t) = \epsilon t$, where $\epsilon$ is a small
positive number \cite{bal2012introduction}.

Stability of the inverse plays a large role in applying the model inverse
to real data or simulated data, which originates from a measurement model with
greater realism.  These possibilities can be conceptually summarized as a
``physical'' measurement model
\begin{equation*}
\mathcal{M}_\text{phys}(x) = \mathcal{M}(x) + \epsilon_\text{det}(x) + \epsilon_\text{noise}(x),
\end{equation*}
where $\epsilon_\text{det}(x)$  and $\epsilon_\text{noise}(x)$ represent, respectively,
deterministic and stochastic error, which are both possibly dependent on $x$.
Applying $\mathcal{M}^{-1}$ to $\mathcal{M}_\text{phys}$, yields
\begin{equation}
\label{physinverse}
\mathcal{M}^{-1} \mathcal{M}_\text{phys}(x) = x + \delta,
\end{equation}
where the magnitude of $\delta$ is bounded with the stability inequality in Eq. (\ref{stability})
with $y_1 = \mathcal{M}(x)$ and $y_2=\mathcal{M}_\text{phys}(x)$. 

With this brief background, there is context for discussing what is meant by solving an
inverse problem. The only step in the analysis of an inverse problem that involves
solving an equation is deriving an inverse for the measurement model
\begin{equation*}
y=\mathcal{M}(x).
\end{equation*}
The model inverse is said to solve the measurement model equation if it can be shown that
\begin{equation}
\label{inverse}
\mathcal{M}^{-1} \left( \mathcal{M} (x) \right) = x.
\end{equation}
Solving an inverse problem should not be confused with applying this model inverse to
a different model that perhaps describes the same physical system, with a greater degree of realism, i.e.
\begin{equation*}
\mathcal{M}^{-1} \left( \mathcal{M}_\text{phys} (x) \right) \neq x.
\end{equation*}
The way that $\mathcal{M}_\text{phys}$ enters the inverse problem analysis
is through the stability inequality in Eq. (\ref{physinverse}).
This inequality in turn relies on 
deriving the inverse to the idealized measurement model $\mathcal{M}$.

\subsection*{Numerical investigation of inverse problems}

Many inverse problems of interest are not amenable to analytic methods and are investigated numerically
through the development of numerical algorithms.
In the numerical setting, the parameter and data spaces are necessarily finite-dimensional
and accordingly $x$ and $y$ are finite length vectors.
A numerical algorithm for inverting $\mathcal{M}$ codes the action of a proposed model inverse.
In numerical investigation of an inverse problem, we have a measurement model and a proposed
inverse to that model.
We may not know: (1) that the measurement
model is injective, (2) that the algorithm implements the proposed model inverse,
or (3) that a proposed inverse actually inverts the measurement model.

The stages of numerical inverse problem investigation starts with verifying that the
algorithm does indeed implement the proposed model inverse.
Having verified the algorithm, it is possible that the proposed model inverse does not actually invert
the measurement model; thus the next step is to generate simulated data
\begin{equation*}
y_\text{test} = \mathcal{M}(x_\text{test})
\end{equation*}
from a realization $x_\text{test}$ and observe that $x_\text{test}$ can be recovered by applying
the algorithm to $y_\text{test}$.
If, however, the verified algorithm fails to recover $x_\text{test}$ there are two possible reasons
for the failure: the proposed model inverse might not invert the measurement model, or the measurement
model itself might not be injective and it is not possible to construct a model inverse at all.
Numerical investigation of the model inverse stability is also important, and
to some extent, accurate recovery of the test image
also provides evidence for stability in that numerical error does not get magnified substantially,
but a thorough investigation on stability entails a characterization of the inverse in response
to data inconsistency and noise.

There are limitations to numerical inverse problem studies. They are empirical by nature.
Knowledge of successful measurement model inversion is limited to
the set of trial model parameter sets $x_\text{test}$ that are successfully recovered.
Furthermore, in most cases of interest, $x_\text{test}$ is only recovered to within some level of
computational error.  One can only amass evidence that a proposed model inverse actually inverts the measurement
model of interest. As part of this evidence it is crucial to report metrics that reflect the verification
of the algorithm and metrics that quantitate the recovery of $x_\text{test}$.

\subsection*{Convex optimization-based inverses}

For CT image reconstruction, there has been much recent work on the use of convex optimization
to provide an inverse to a relevant measurement model. In this setting, the model inverse
is formally expressed as
\begin{align*}
x^\star=&\mathcal{M}^{-1}(y_\text{test}) \\
=& \argmin_x \Phi[y_\text{test}](x)
\text{ such that } x \in S_\text{convex}[y_\text{test}]
\end{align*}
where $\Phi$ is a convex objective function and $x$ is restricted to convex
set $S_\text{convex}$, encoding prior knowledge on $x$. Either $\Phi$ or $S_\text{convex}$
may depend on data vector $y_\text{test}$. In the vast majority of cases of interest to
CT image reconstruction, this optimization problem is solved by an iterative algorithm.

Numerical algorithm verification consists of (1) showing that the iterates approach the optimality
conditions of this optimization problem and (2) showing that the iterates generate data that
approaches $y_\text{test}$. Approaching the optimality conditions ensures that
\begin{equation}
\label{xapproach}
x_k \rightarrow x^\star \text{ as } k \rightarrow \infty,
\end{equation}
where $k$ is the iteration index of the algorithm.
Approaching the test data
\begin{equation*}
\mathcal{M}(x_k) \rightarrow y_\text{test}
\end{equation*}
can be measured by the data root-mean-square-error (RMSE)
\begin{equation}
\label{dapproach}
\sqrt{\|y_\text{test} -\mathcal{M}(x_k)\|^2_2/\text{size}(y_\text{test})} \rightarrow 0.
\end{equation}
Showing evidence for Eqs. (\ref{xapproach}) and (\ref{dapproach}) verifies that the algorithm
implements the proposed inverse.

Evidence that the proposed inverse actually inverts the measurement model is established
by showing that the iterates
of the numerical algorithm approaches test image
\begin{equation*}
x_k \rightarrow x_\text{test}.
\end{equation*}
This can be measured by the image root-mean-square-error (RMSE)
\begin{equation}
\label{iapproach}
\sqrt{\|x_\text{test} -x_k\|^2_2/\text{size}(x_\text{test})} \rightarrow 0.
\end{equation}
We reiterate that a failure to show Eq. (\ref{iapproach}) can either be a result of the
non-injectivity of the measurement model -- i.e. there is more than one image that yields
$y_\text{test}$ -- or the proposed inverse does not invert $\mathcal{M}$.

\section{Sparse-view CT and compressed sensing}
\label{sec:sparseview}

The most common measurement model for the CT system involves line integration over an object
model. The line-integration model takes the form of the X-ray or Radon transform
for a divergent ray or parallel ray geometry, respectively. The present study employs a parallel beam
geometry in a 2D CT scanning setting. For iterative image reconstruction, the continuous line-integration
model is converted to a discrete-to-discrete transform by representing the object function with
an image array of pixels.
The measurement model parameters are the
values of the image pixels and the measurement model $\mathcal{M}$ is
\begin{equation}
\label{RadonModel}
g= R f,
\end{equation}
where $f$ denotes the pixel values; $R$ is a discrete-to-discrete linear
transform representing the action
of the Radon transform on the image pixels; and $g$ indicates the sinogram values.
Analyzing Eq. (\ref{RadonModel}) as an inverse problem is straight-forward. If the matrix $R$
is left-invertible then the measurement model is one-to-one and two different sets of pixel
values, $f$, yield different sinograms $g$ and a left-inverse $R^{-1}$ can be constructed.
Stability of performing this inversion depends on the singular value
spectrum of the particular discrete-to-discrete version of the Radon transform \cite{jakob2012quantifying}.

For sparse-view sampling, the number of projections in the sinogram is small enough that $R$ is a short,
fat matrix; i.e. there are more columns than rows. In this case, injectivity is lost and it can occur that
two different images $f$ can yield the same sinogram under the action of $R$.
The conjecture of Compressed Sensing \cite{candes2006robust,donoho2006compressed} is that injectivity
of Eq. (\ref{RadonModel}) can be restored by restricting the image $f$ to be sparse
or sparse in some transform $Tf$.

The modified measurement model $\mathcal{M}$ studied in Compressed Sensing is thus
\begin{equation}
\label{RadonModelSparseTransform}
g= R f \text{ where } f \in F_\text{sparse}
\text{ and } F_\text{sparse} = \{f \; | \; \|Tf\|_0 \le s\},
\end{equation}
where $T$ is the sparsifying transform;
$\| \cdot \|_0$ is the counting norm, which yields the number of non-zero elements in the argument
vector; and $s$ is an integer bounding the number of non-zeros in $Tf$. Direct restriction in the sparsity
of $f$ is the special case where $T=I$, the identity matrix. Note that the 
the measurement model $\mathcal{M}$ includes both the transform $R$ and the sparsity restriction on $f$.

The canonical inverse to this measurement
model is specified implicitly by the formal optimization problem
\begin{equation}
\label{L0min}
f^\star = \argmin_f \|Tf\|_0 \text{ such that } g = Rf.
\end{equation}
From a computational point of view this inverse is not practical to evaluate, especially
for full-scale CT systems, due to the $\ell_0$-norm in the objective function.

As suggested in Ref. \cite{candes2006robust}, a more practical inverse
to the sparsity-restricted measurement model
may be obtained by
the convex relaxation of the $\ell_0$-norm in Eq. (\ref{L0min})
to the $\ell_1$-norm.
\begin{equation}
\label{L1min}
f^\star = \argmin_f \|Tf\|_1 \text{ such that } g = Rf.
\end{equation}
where $\| \cdot \|_1$ is the $\ell_1$-norm, which sums the absolute value of the vector components.
While analytic results for Compressed Sensing have been established for certain classes of measurement
models \cite{candes2008introduction}, these results do not cover the line-integration
based measurement models used in CT and we must turn to numerical investigation and amass results
from CT simulations.

With numerical simulation, the algorithm
is verified by showing that the solution of Eq. (\ref{L1min}) is attained.
Successful recovery of test image $f_\text{test}$ from data $g_\text{test}=Rf_\text{test}$ is evidence
that Eq. (\ref{L1min}) provides the inverse of
Eq. (\ref{RadonModelSparseTransform}) and that measurement
model Eq. (\ref{RadonModelSparseTransform}) is one-to-one. Failure to recover $f_\text{test}$ could mean either that
the measurement model is not injective or that the proposed inverse does not actually invert the measurement model.
The convex relaxation of the $\ell_0$-norm to the $\ell_1$-norm, in particular, does raise the distinct possibility
that the Compressed Sensing model could be injective while Eq. (\ref{L1min}) might not provide its inverse.

For Compressed Sensing style studies
the goal is to establish a relationship between the number of samples needed for accurate image
recovery and sparsity in the image or transformed image. A practical approach to achieving this
relationship numerically is to perform a type of Monte Carlo investigation where test images are generated
randomly at various sparsity levels. The information on the sampling level required to recover the test
images as a function of their sparsity can be summarized in a Donoho-Tanner
phase diagram \cite{donoho2005sparse,jorgensen2015little}.
For full-scale CT systems it is of interest to perform numerical inverse
problem related investigations on configurations and test objects that reflect properties of a CT system
of interest. We present such studies here so as to have a reference for numerical inverse problem
studies with CNNs.

\subsection*{Gradient sparsity exploiting image reconstruction}
\label{sec:TVmin}

A particularly successful sparsity model for CT imaging is seeking sparsity in the image gradient magnitude,
because most of the variation in medical CT images occurs at the organ or tissue boundaries \cite{Sidky2006,sidky2008image}.
For this type of sparsity the most common sparsifying transform is
\begin{equation*}
Tf = |Df|_\text{mag},
\end{equation*} 
where $D$ is the finite differencing approximation to the image gradient, and $|\cdot|_\text{mag}$
represents the spatial vector magnitude; i.e. $Df$ is the gradient of $f$ and $|Df|_\text{mag}$ is the
gradient magnitude image (GMI).

\begin{figure}[!t]
\centerline{\includegraphics[width=0.95\columnwidth]{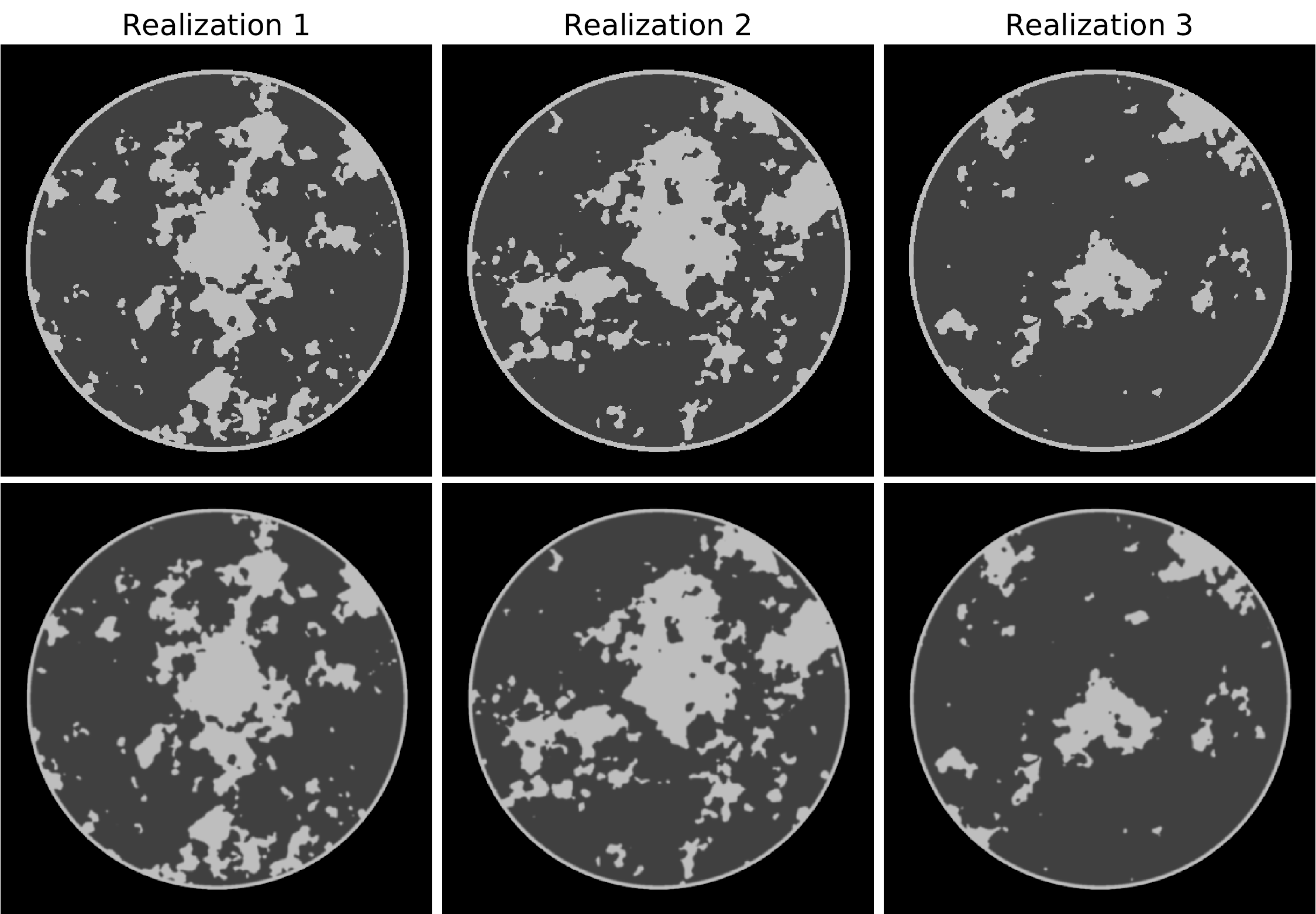}}
\caption{Three realizations of a 
computerized breast phantom with stochastic distribution of fibroglandular tissue.
The top row shows realizations of the binary phantom, and the bottom row shows
the corresponding smooth-edge phantom, generated by blurring with a gaussian
kernel of one pixel width.
The breast phantom is composed of a 16 cm disk containing 
background fat tissue, attenuation 0.194 cm$^{-1}$, skin-line and 
fibroglandular tissue at attenuation 0.233 cm$^{-1}$.
The generated phantom images are 512 $\times$ 512 pixels and are
shown in a gray scale window of [0.18, 0.24] cm$^{-1}$.}
\label{fig:phantom}
\end{figure}

The measurement model $\mathcal{M}$ is
\begin{multline}
\label{SparseGMI}
g= R f \text{ where } f \in F_\text{GMI-sparse}\\
\text{ and } F_\text{GMI-sparse} = \{f \; | \; \|(|Df|_\text{mag})\|_0 \le s\},
\end{multline}
where again the measurement model includes the restriction that $f \in F_\text{GMI-sparse}$.
The posited inverse of this measurement model is defined implicitly with the optimization problem
\begin{equation}
\label{TVmin}
f^\star = \argmin_f \|(|Df|_\text{mag})\|_1 \text{ such that } g = R f.
\end{equation}
where $\|(|Df|_\text{mag})\|_1$ is also known as the image total variation (TV).
We refer to this optimization problem as equality constrained TV minimization (TVmin).
Numerical investigations are carried out by generating simulation data according to Eq. (\ref{SparseGMI})
and numerically solving Eq. (\ref{TVmin}) to see if the test phantom is recovered accurately.
The TVmin problem can be efficiently solved by the Chambolle-Pock primal-dual (CPPD)
algorithm \cite{chambolle2011first,Pock2011,SidkyCP:12}
explained in Appendix \ref{app:cppd} with the pseudo-code listed in Algorithm \ref{alg:cppd}.

Accurate recovery of the test phantom provides evidence for injectivity of the measurement model,
because it is unlikely that the original GMI-sparse test image would be recovered if there were other
GMI-sparse images that yield the same data.  Of course, recovery of the test image also provides
evidence that TVmin inverts Eq. (\ref{SparseGMI}). Also, stability is tested to the
extent that numerical error does not get amplified and interfere with accurate recovery
of the test phantom.

\begin{figure}[!t]
\centerline{\includegraphics[width=0.95\columnwidth]{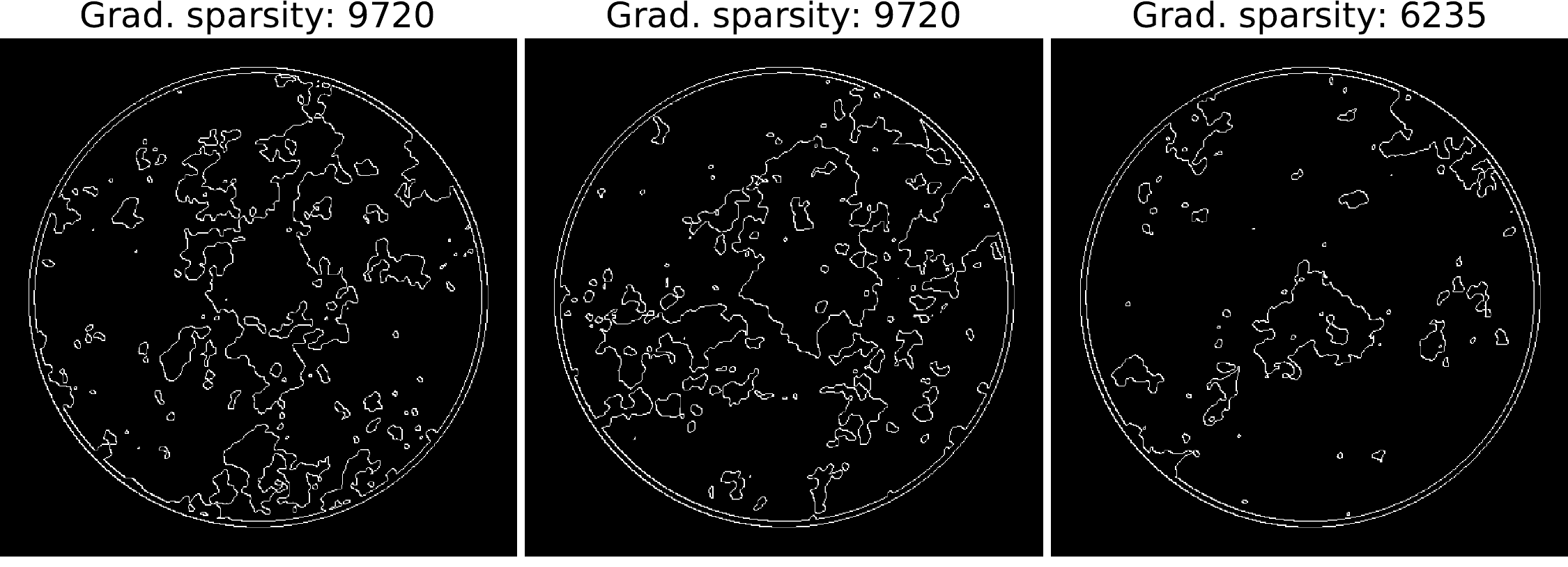}}
\caption{Non-zero pixels, indicated in white,
of the gradient magnitude images (GMI) corresponding to the
phantoms shown in Fig. \ref{fig:phantom}. The GMI sparsity is listed above
each panel. For reference, the total number of pixels in the image array
is $512^2=262,144$. Thus the corresponding percentage of GMI non-zero
pixels is 3.7\%, 3.7\%, and 2.37\% from left to right.}
\label{fig:gmi1}
\end{figure}

\section{CNN-based deep-learning for sparse-view CT}
\label{sec:CNN}

We implement CNN-based deep-learning for image reconstruction for a 2D sparse-view problem, where
a 512$\times$512 image is reconstructed from a sinogram of dimension 128 views by 512 detector bins.
In order to implement CNN image reconstruction, 4,000 simulation phantoms are generated from
the stochastic models described in Sec. \ref{sec:phantom}. From each phantom, a 128-view sinogram
is generated by use of Eq. (\ref{RadonModel}). The 128-view sinograms are further processed by applying
filtered back-projection
(FBP). The resulting FBP-reconstructed images,
which have prominent streak artifacts due to the sparse-view sampling,
are paired with the ground truth images and used to train the CNN.
The data available to training, validating, and testing the CNN thus consists of 4,000 paired images,
where each pair consists of the true phantom and its 128-view FBP reconstructed counterpart.
The goal of the CNN training is to obtain a model that will yield the true phantom from the FBP reconstruction.

The CNN uses a standard U-net architecture as described in Figure 4a of Ref.  \cite{han2018framing}.
The output of the U-net estimates the difference between the 128-view FBP images and its phantom
counterpart. Given a 128-view FBP image, the CNN-predicted residual is subtracted from the
the FBP input to obtain the predicted reconstruction from the 128-view data.

For each dataset of 4000 pairs of images, 3990 are used for training and validation, and 10 images
are left out for independent testing. Of the 3990 pairs, 80\% or 3192 are used for training the
residual U-net and 20\% or 798 are used for validation. The models are trained by minimizing the
mean-square-error (MSE) between the residual labels and predictions using a stochastic gradient descent algorithm (SGD)
with momentum \cite{rumelhart1986learning}, learning rate decay and
Xavier weight initialization \cite{glorot2010understanding}.
Model training is done for 450 epochs using the MATLAB MatConvNet toolbox \cite{vedaldi2015matconvnet}
and a NVIDIA GeForce GTX 1080 Ti graphics processing unit (GPU) for computation.
For the present computations, the training took one week to complete.

\begin{figure}[!t]
\centerline{\includegraphics[width=0.95\columnwidth]{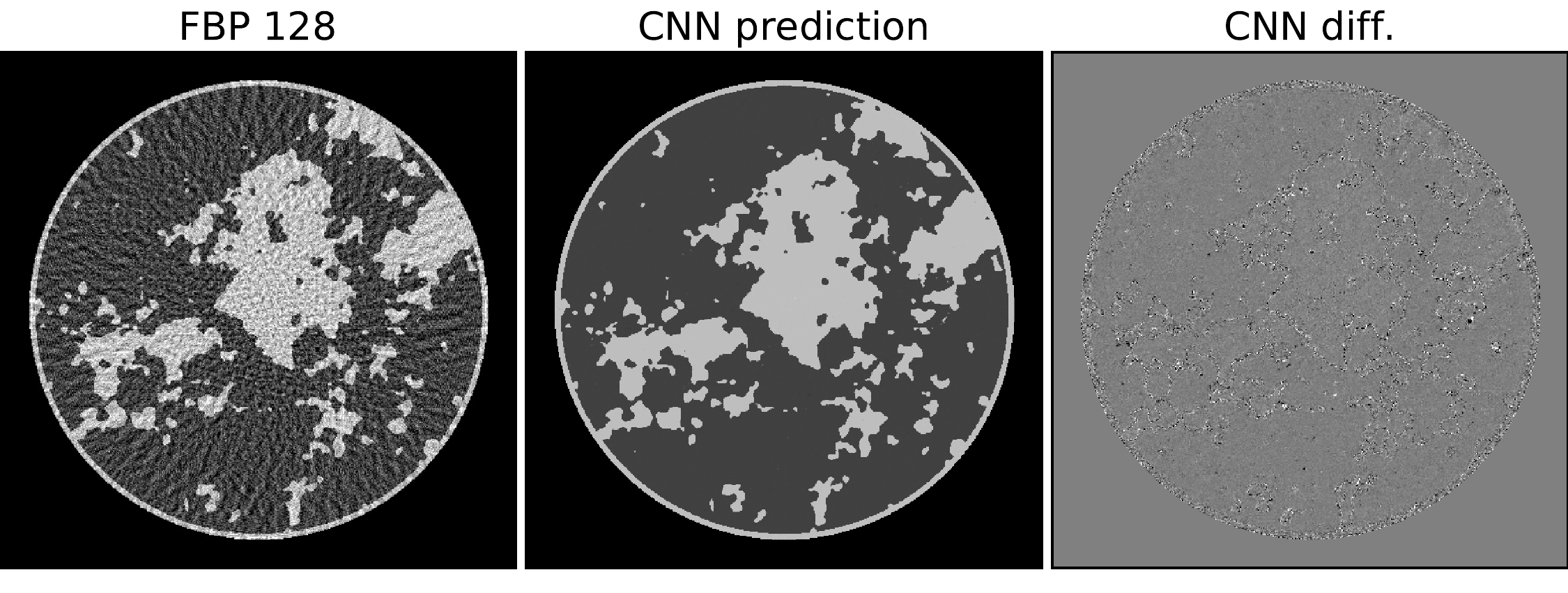}}
\centerline{\includegraphics[width=0.95\columnwidth]{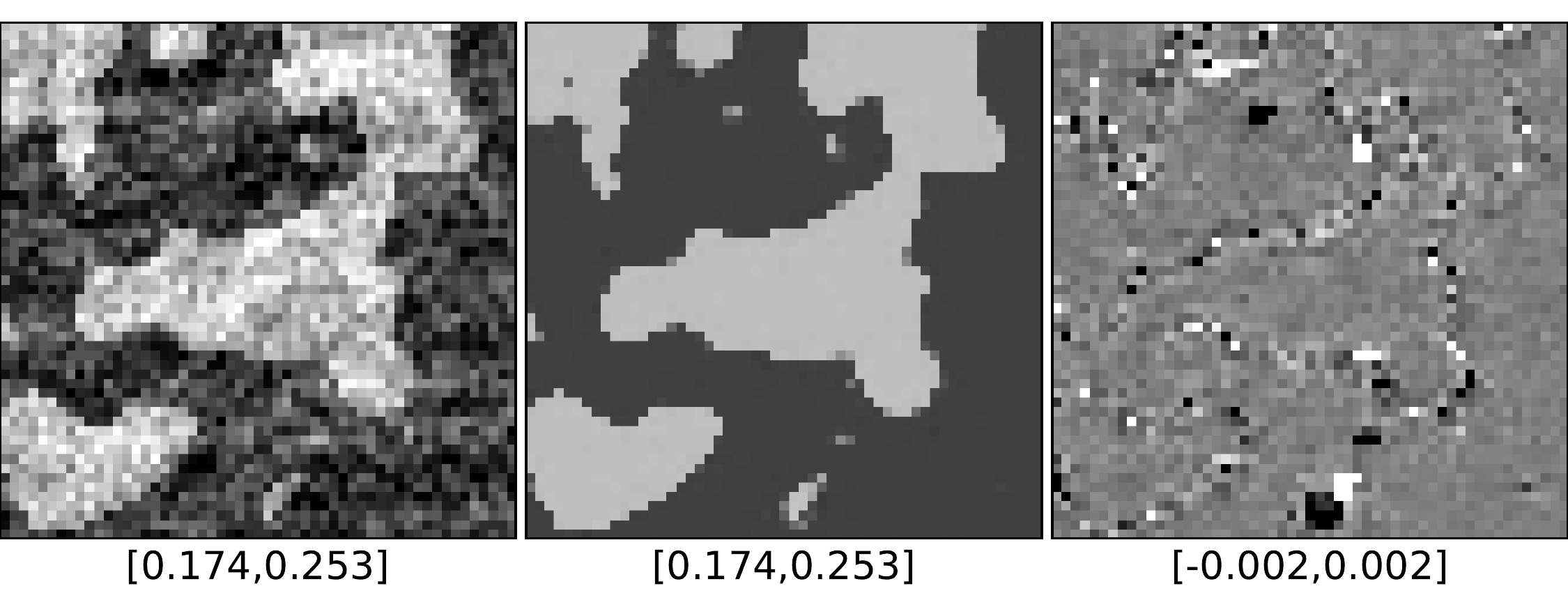}}
\caption{Results for the CNN on reconstruction of a binary test phantom.
Shown are FBP applied to the 128-view sinogram that serves as input
to the CNN, the CNN reconstruction, and the corresponding difference images.
The whole images appear in the top row and a corresponding ROI image is displayed
in the bottom row.
The gray scale windows are specified below the ROI panels in units of cm$^{-1}$.
The RMSE for the CNN result is $1.15\times10^{-3}$ cm$^{-1}$ and
there is obvious discrepancy in the difference image where the
shown grayscale window is 10\% of the contrast between fat and fibroglandular tissue.
The CNN image error is highly non-uniform and the maximum pixel error is $0.04$ cm$^{-1}$,
which is the same as the phantom tissue contrast.}
\label{fig:binaryCNN}
\end{figure}

One of the challenging aspects of analyzing deep-learning in the framework of inverse
problems is that the measurement model itself is not completely specified.
For the sparse-view CT setting,
we do know that the data are generated by the discrete-to-discrete Radon transform,
but the restriction on the images
in the domain of the transform is not described mathematically. Instead there is the
vague notion that the image domain is a set $S$ that is somehow defined by
the set of training data $F_\text{train}$
\begin{equation}
\label{RadonModelCNN}
g= R f \text{ where } f \in S(F_\text{train}).
\end{equation}
While it is not clear from the literature how to define $S$ in the CT setting, $S$
must include at least the training data if CNNs are truly capable of inverting
Eq. (\ref{RadonModelCNN}). Thus, in the absence of a completely specified measurement model
we can still check that a training image is recovered if the CNN is given the FBP
reconstruction of data corresponding to that training image.

Because we do not know what the precise measurement model is for deep-learning, we employ
a measurement model that is based on gradient-sparsity restriction. Knowledge of the measurement
model is made available to the CNN by generating training and testing image/data pairs
that adhere perfectly to the idealized gradient-sparsity restricted model.
We test the deep-learning methodology on its ability to learn and invert this measurement model.

\section{Breast phantom realizations}
\label{sec:phantom}

\begin{figure}[!t]
\centerline{\includegraphics[width=0.95\columnwidth]{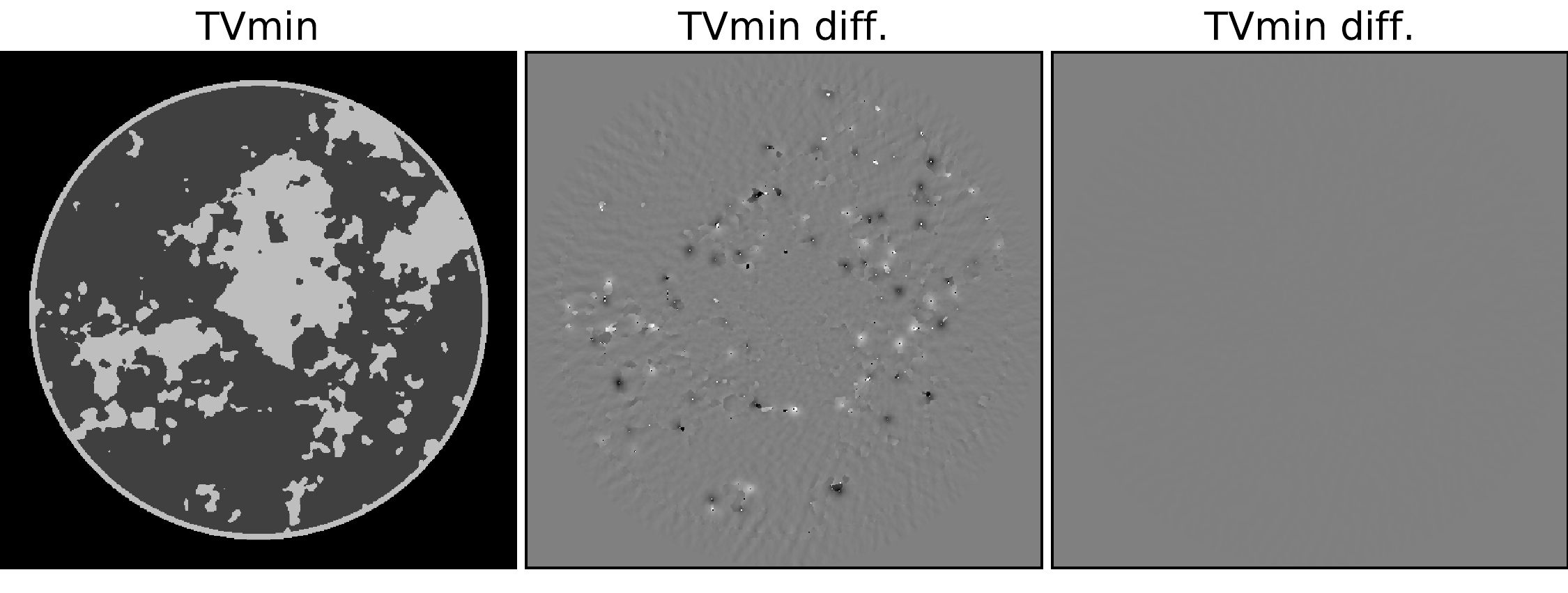}}
\centerline{\includegraphics[width=0.95\columnwidth]{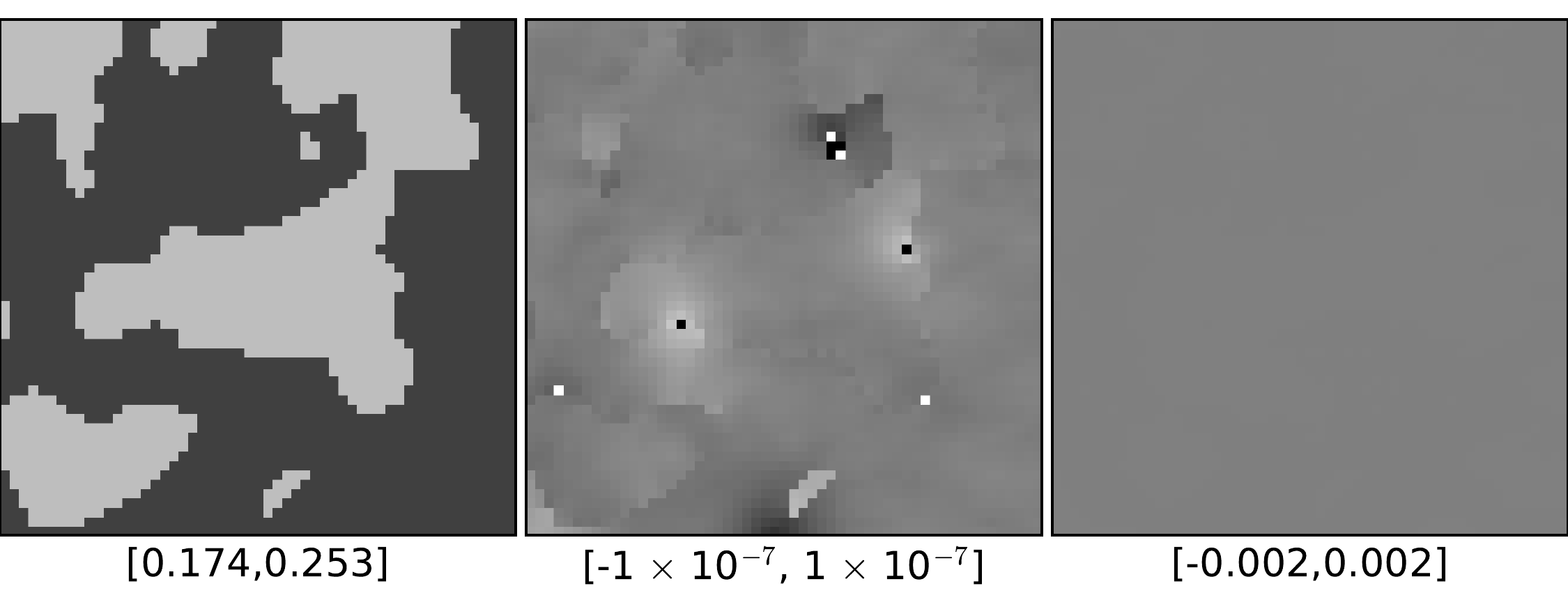}}
\caption{Results for TVmin on reconstruction of a binary test phantom.
Shown are the TVmin reconstruction and the corresponding difference image in two
different gray scale windows. The narrow gray scale range in the middle column
is selected so that the difference from the truth can be seen, and the gray scale
window in the right column is the same as that of the corresponding CNN result.
The whole images appear in the top row and a corresponding ROI image is displayed
in the bottom row.
The gray scale windows are specified below the ROI panels in units of cm$^{-1}$.
The RMSE for the TVmin result is $6.43\times10^{-8}$ cm$^{-1}$ and
the maximum deviation from the truth over all pixels is $7.11\times10^{-6}$ cm$^{-1}$.}
\label{fig:binaryTVmin}
\end{figure}

The breast CT slice phantom is designed to provide an illustrative test for both
the gradient sparsity exploiting TVmin algorithm and CNN-based deep-learning.
The phantom consists of modeled fat, fibroglandular,
and skin tissues. The distribution of the fibroglandular tissue is generated
by thresholding a power-law noise realization described in Ref. \cite{Reiser10}.
The phantom images are generated on a 512$\times$512 pixel grid.

From the basic phantom realizations, two image classes are defined.
The first class, called ``binary'', consist of the image realizations themselves,
and the transition between fat and fibroglandular tissues is sharp.
The second class, called ``smooth-edge'', is derived from the binary class
by convolving 
with a gaussian kernel where the full width at half maximum (FWHM) is set to
one pixel
\begin{equation*}
f_\text{smooth-edge} = G(w_0) f_\text{binary} \text{ and } w_0 =1,
\end{equation*}
where $G(w)$ symbolizes the operation of gaussian convolution with width parameter $w$.
Three realizations of the ``binary'' and ``smooth-edge'' phantoms are shown in
Fig. \ref{fig:phantom}. Having the two classes of images allow us to test TVmin and
the CNN under each of the two classes of images and under a generalized object
model where the images can be selected from either class.

The binary phantom realizations have a sparse gradient magnitude image (GMI) as
shown in Fig. \ref{fig:gmi1}, where the three shown realizations all indicate
a high degree of GMI sparsity. The maximum number of GMI non-zeros over all 4000 phantom
realizations is 12053, which corresponds to 4.58\% of the whole image.
Sparse GMI should allow for accurate image reconstruction
by TVmin with significantly reduced sampling even though the tissue borders have
a complex geometry.

The stochastic nature of the phantom makes it ideal for testing data-driven
image reconstruction because the number of phantom realizations
is only limited by storage space and computation time. Furthermore,
the image truth is known exactly, eliminating error due to imperfect labeling of the training data.

\begin{figure}[!t]
\centerline{\includegraphics[width=0.95\columnwidth]{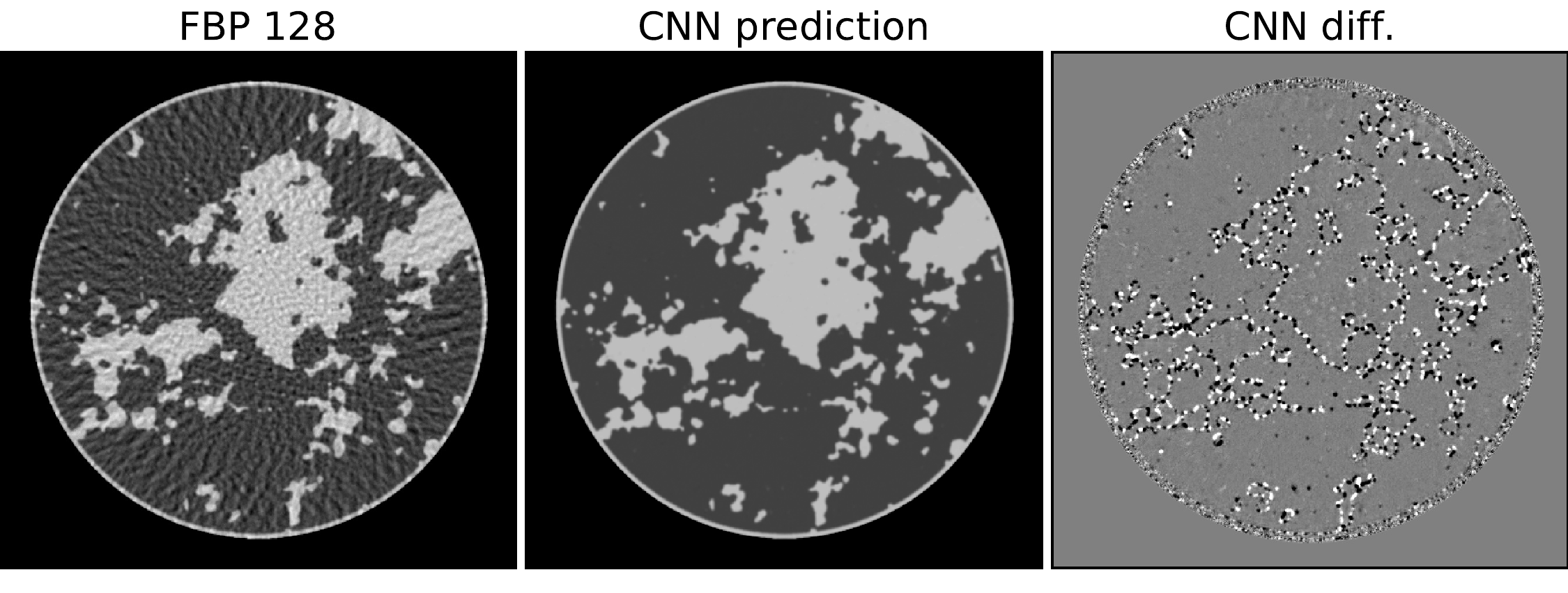}}
\centerline{\includegraphics[width=0.95\columnwidth]{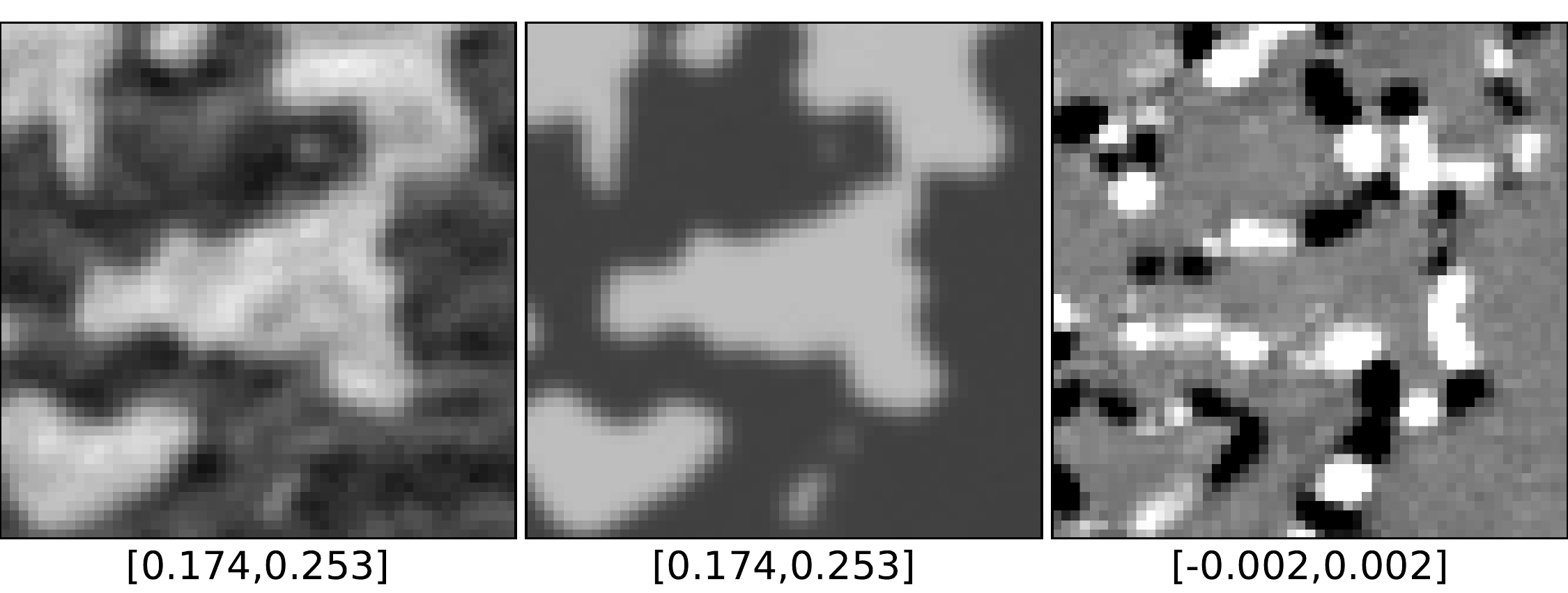}}
\caption{Results for the CNN on reconstruction of a smooth-edge test phantom.
The format of the figure is the same as that of Fig. \ref{fig:binaryCNN}.
The RMSE for the CNN result is $6.76\times10^{-4}$ cm$^{-1}$ and
the maximum pixel error is $1.52 \times 10^{-2}$ cm$^{-1}$.}
\label{fig:smoothedgeCNN}
\end{figure}

\section{Results}
\label{sec:results}
For the sparse-view CT problem of study, we generate parallel-beam projections of the
binary or smooth-edge breast phantom, where $R=R_{128}$ is specified as a circular scan over
360 degrees with 128 projections. The linear detector array has length 18 cm
and consists of 512 bins. The image array is also 18$\times$18 cm$^2$ consisting
of 512x512 pixels, the same as the image array used for the breast phantom.
With this sampling scheme, it is clear that the measurement model
\begin{equation*}
g=R_{128}f
\end{equation*}
without any restriction on $f$
is not injective because the size of $g$ is 128$\times$512 measurements and the number
of unknowns pixel values is 512$\times$512. There are four times as many unknowns as knowns,
and accordingly $R$ has a non-trivial nullspace meaning that there can be two images
that yield the same data on applying $R_{128}$.

For gradient-sparsity exploiting image reconstruction,
the additional sparsity restriction on the image possibly allows
the measurement model to be inverted.
For deep-learning
with CNNs, the training data are exploited to ``learn'' the measurement model which
should include learning the subset $S$ to which the object model images are restricted.

\begin{figure}[!t]
\centerline{\includegraphics[width=0.95\columnwidth]{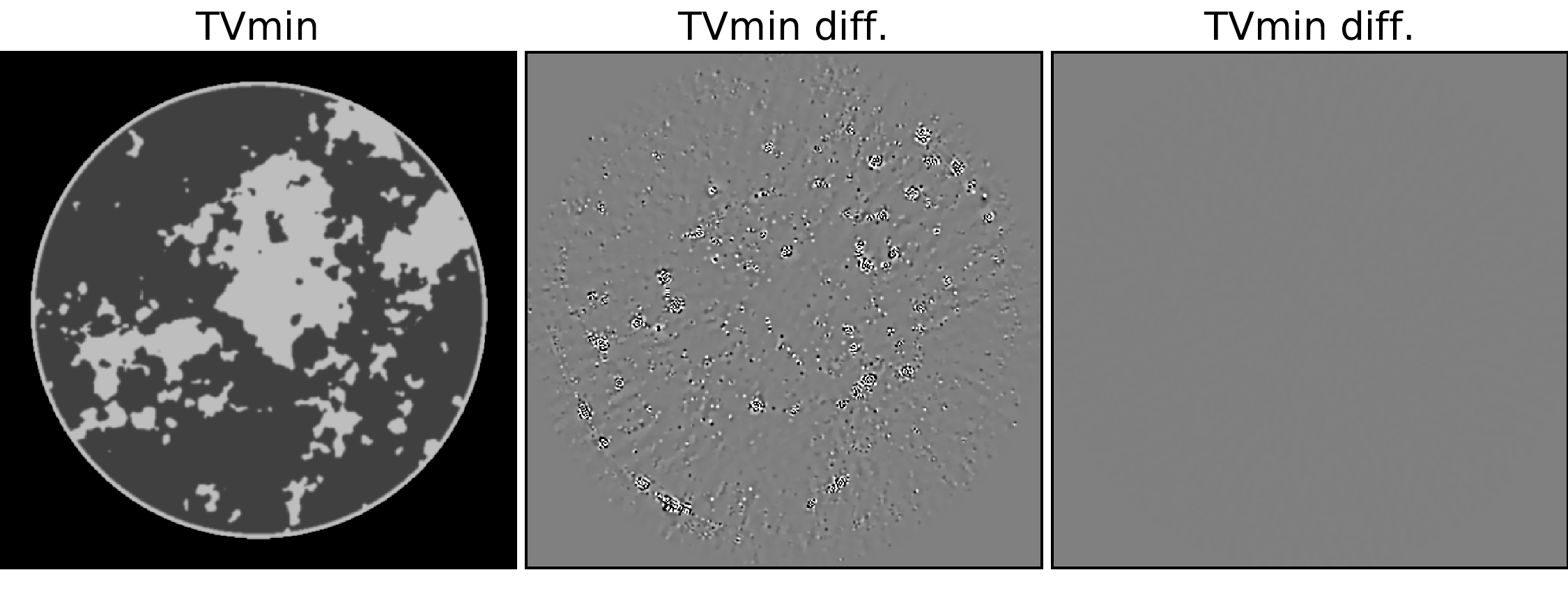}}
\centerline{\includegraphics[width=0.95\columnwidth]{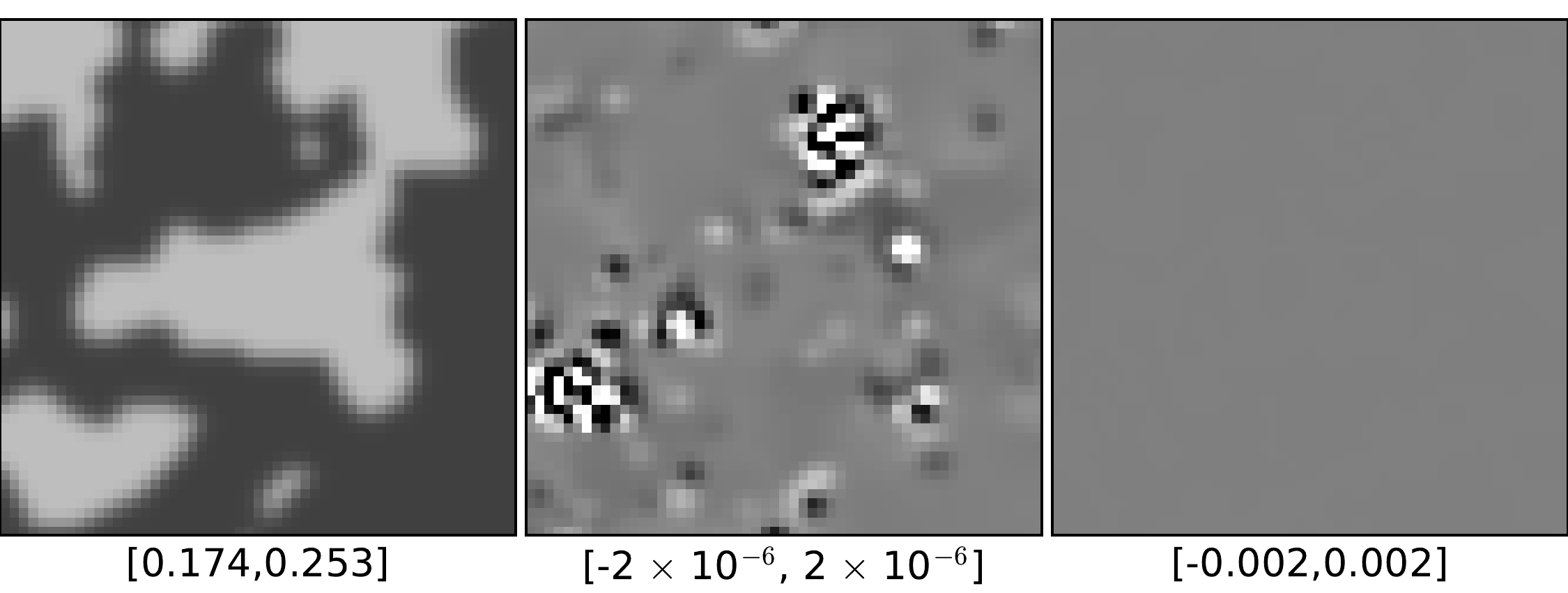}}
\caption{Results for TVmin on reconstruction of a smooth-edge test phantom.
The format of the figure is the same as that of Fig. \ref{fig:binaryTVmin}.
The RMSE for the TVmin result is $1.15\times10^{-6}$ cm$^{-1}$ and
the maximum deviation from the truth over all pixels is $7.64\times10^{-5}$ cm$^{-1}$.}
\label{fig:smoothedgeTVmin}
\end{figure}

\subsection{Image reconstruction with the binary class}
\label{sec:binaryres}
The images, which are drawn from the binary breast phantom class,
have GMI sparsity and consequently the data
are generated from an instance of the  measurement model in Eq. (\ref{SparseGMI})
\begin{multline}
\label{model1}
g= R_{128} f_\text{obj} \text{ where }  f_\text{obj} \in F_\text{binary}\\
\text{ and } F_\text{binary} = \{f \; | \; \|(|Df|_\text{mag})\|_0 \le 12053 \},
\end{multline}
where 12053 is the largest number of GMI non-zeros in the 4000 binary phantom image realizations.
To demonstrate image reconstruction from the binary phantom,
TVmin and the deep-learning CNN are applied to the 128-view sinogram generated from the
second phantom shown in Fig. \ref{fig:phantom}.

The CNN results are shown in Fig. \ref{fig:binaryCNN}, where the input FBP, reconstructed, and
difference images are shown. The top row shows the full reconstructed image array, and the bottom
row shows a corresponding region of interest (ROI).
While the FBP image looks as if it is generated from noisy
data, it is actually not. The complexity of the edge structure and the severe angular
undersampling causes the wavy and noisy appearance. 

On the global and ROI views the CNN prediction images appear to be accurate in
the shown gray scale window.
Some discrepancies are visible in the ROI image.
The difference images, however, reveal substantial
discrepancy between the CNN reconstruction and the true phantom image in a gray scale
window that is 10\% of the fat/fibroglandular contrast. Furthermore, the maximum pixel error
is at the level of this soft-tissue contrast. As a result, the claim that the CNN provides
a numerically accurate reconstruction and numerically solves Eq. (\ref{model1}) can not be made.

\begin{figure*}[!t]
\centerline{\includegraphics[width=0.95\columnwidth]{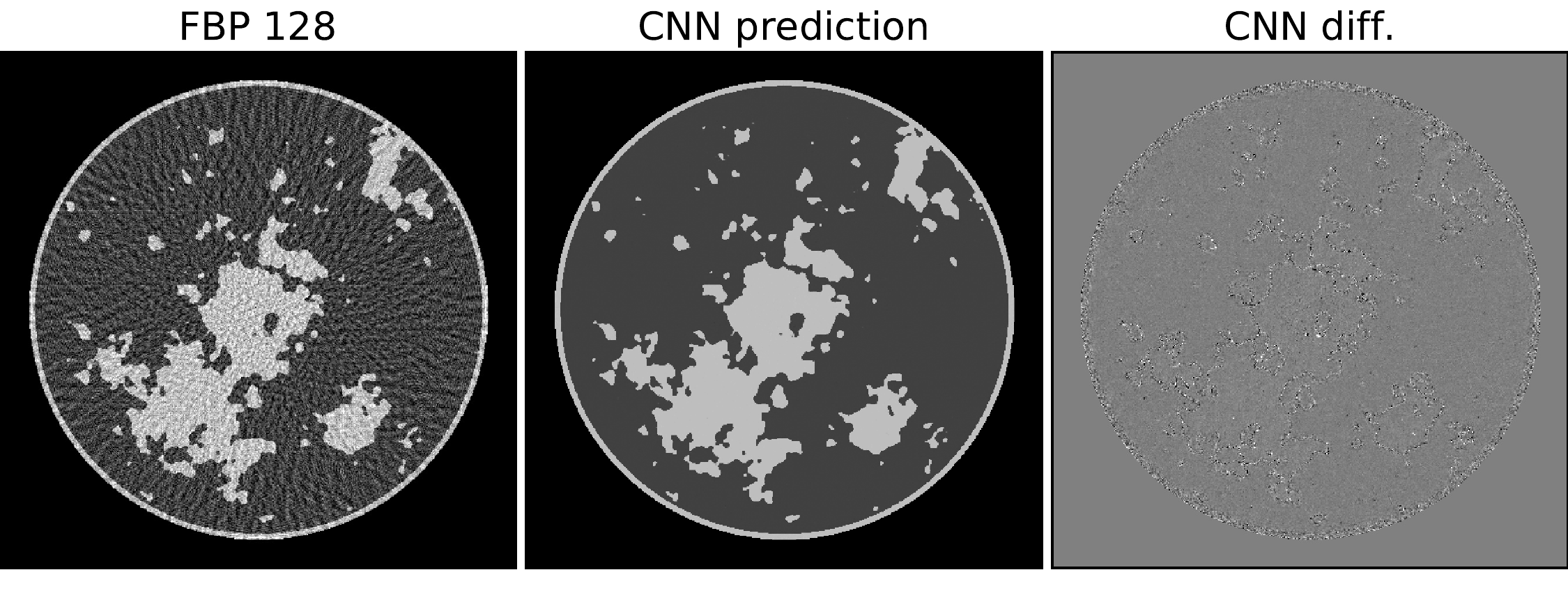}~~~
            \includegraphics[width=0.95\columnwidth]{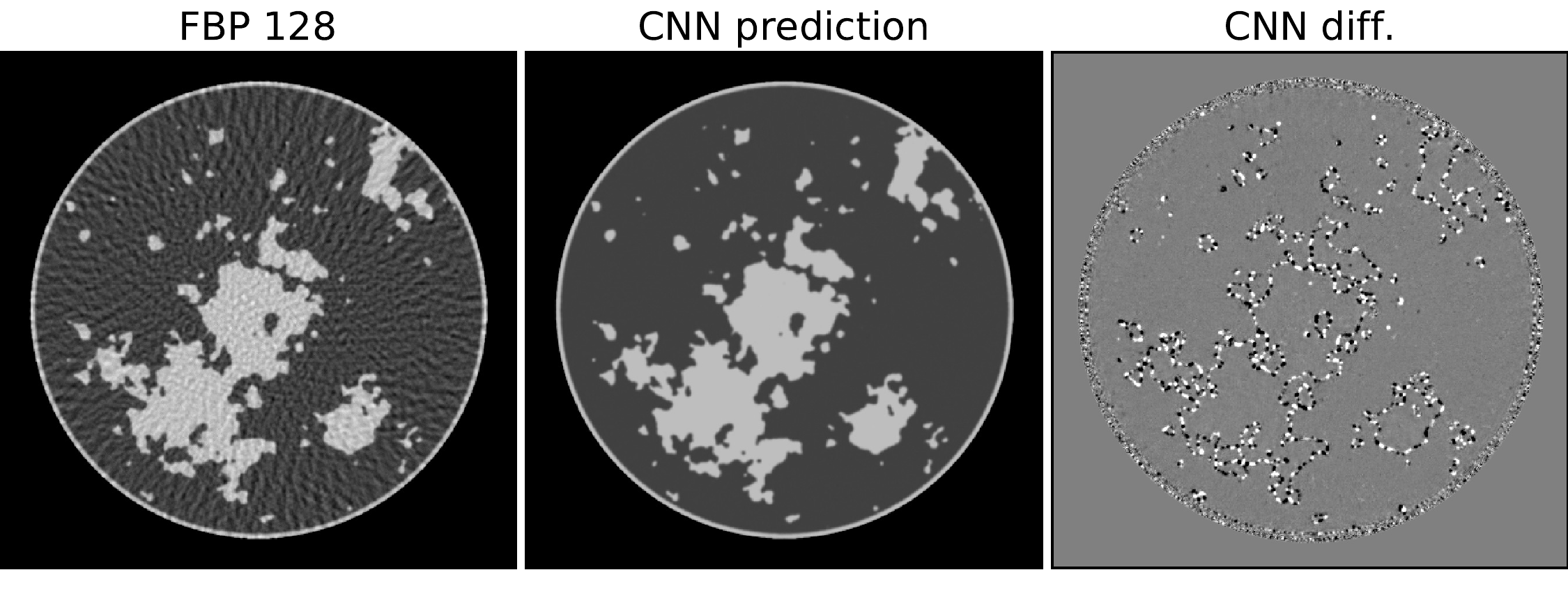}}
\centerline{\includegraphics[width=0.95\columnwidth]{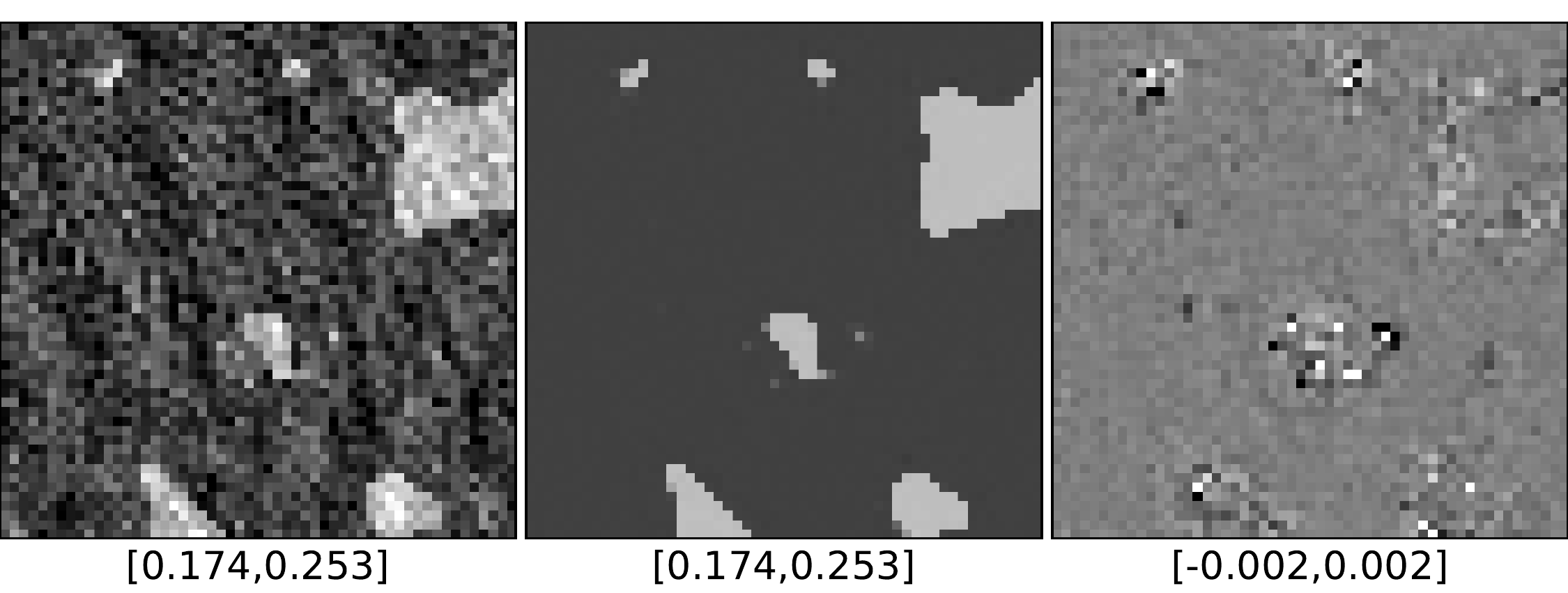}~~~
            \includegraphics[width=0.95\columnwidth]{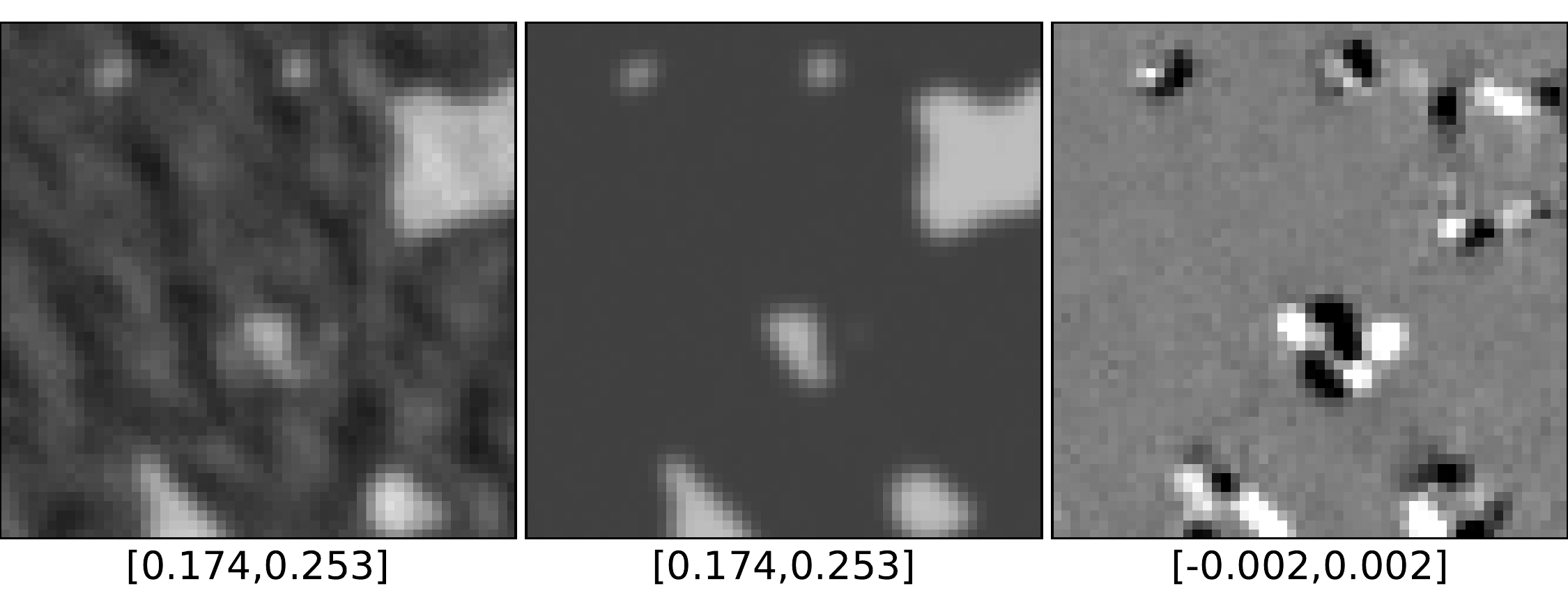}}
\caption{The left and right sets of results follow the same format as Figs. \ref{fig:binaryCNN}
and \ref{fig:smoothedgeCNN}, respectively.
The respective CNNs are applied to images used in the training.
For the binary training image on the left, the RMSE is $7.90\times10^{-4}$ cm$^{-1}$ and
the maximum pixel error is $0.04$ cm$^{-1}$.
For the smooth-edge training image on the right,
the RMSE is $4.66\times10^{-4}$ cm$^{-1}$ and
the maximum pixel error is $0.011$ cm$^{-1}$.}
\label{fig:inTS}
\end{figure*}

The corresponding results for TVmin are shown in Fig. \ref{fig:binaryTVmin},
where it is clear that the numerical accuracy of TVmin is much greater than that of the CNN.
The image RMSE is five orders of magnitude smaller than the soft-tissue contrast, and the maximum pixel
error is three orders of magnitude smaller than this contrast level. As a result, there is no
perceptible difference between the reconstructed image and true phantom for any ROI in the image.
To emphasize the difference in the accuracy of the recovery, the last column of Fig. \ref{fig:binaryTVmin}
displays the TVmin difference images in the gray scale used for the CNN difference images.

The conclusion from this particular experiment is that TVmin is able to numerically
recover the test phantom. This result is not too surprising because the number of non-zero
pixels in the GMI of the phantom
is 9720 pixels and the total number of samples is $128\times512=65536$ or approximately
6.5 times the GMI sparsity.
The CNN, on the other hand, does not
provide a numerically accurate inverse. Furthermore, the CNN is not able to
recover the test phantom in a situation where we {\it know} there exists an accurate
numerical inverse as provided by TVmin.

\subsection{Image reconstruction with the smooth-edge class}
\label{sec:smoothedgeres}

The smooth-edge breast phantom class images have GMI sparsity in the object model and
the corresponding restricted measurement model is
\begin{multline}
\label{model2}
g= R_{128} f_\text{obj} \text{ where }  f_\text{obj} \in F_\text{smooth-edge}\\
\text{ and } F_\text{smooth-edge} = \{G(w_0)f \; | \; \|(|Df|_\text{mag})\|_0 \le 12053 \},
\end{multline}
and $w_0=1$ in pixel units.
In terms of methodology, there is no change in implementing the training of the CNN
for this class of images other than switching to the set of smooth-edge training pairs.
To perform sparsity-exploiting image reconstruction for this measurement model, the optimization
problem in Eq. (\ref{TVmin}) is altered slightly in the data equality constraint, becoming
\begin{equation}
\label{TVmin128}
f^\star = \argmin_f \|(|Df|_\text{mag})\|_1 \text{ such that } g = R_\text{128} G(w_0) f.
\end{equation}
The reconstructed image is obtained from $f^\star$ by applying the Gaussian blurring
\begin{equation*}
f_\text{recon} = G(w_0) f^\star.
\end{equation*}
Generalizing the implementation of TVmin to account for the blur is straight-forward \cite{wolf2013few,zhang2018optimization},
and the necessary changes are discussed in Appendix \ref{app:cppd}.

The results for image reconstruction by the CNN and TVmin are shown in Figs. \ref{fig:smoothedgeCNN}
and \ref{fig:smoothedgeTVmin}, respectively.
Again, both reconstructed images are visually accurate on the global view and in the chosen ROI. For the CNN
results,
the difference images reveal a wider band of discrepancy at the transition between fat and fibroglandular
tissue than was the case for the binary class study. There is a corresponding distortion in the shapes
of the fibroglandular regions that is visible in the ROI.
The numerical results are similar to that of the binary class study. There is some numerical
differences in that the smooth-edge RMSE values are lower than that of the binary phantom case,
but they are still far from being able to claim numerically
accurate image reconstruction.

\begin{figure*}[!t]
\centerline{\includegraphics[width=0.95\columnwidth]{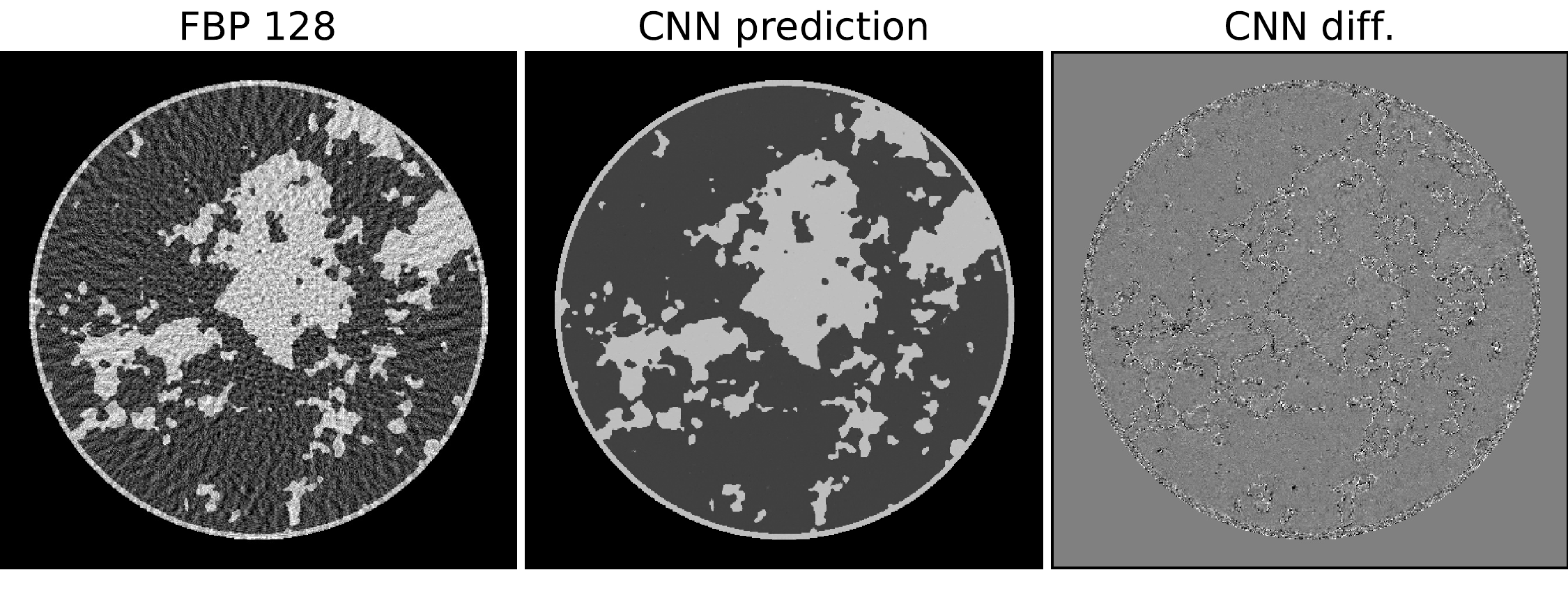}~~~
            \includegraphics[width=0.95\columnwidth]{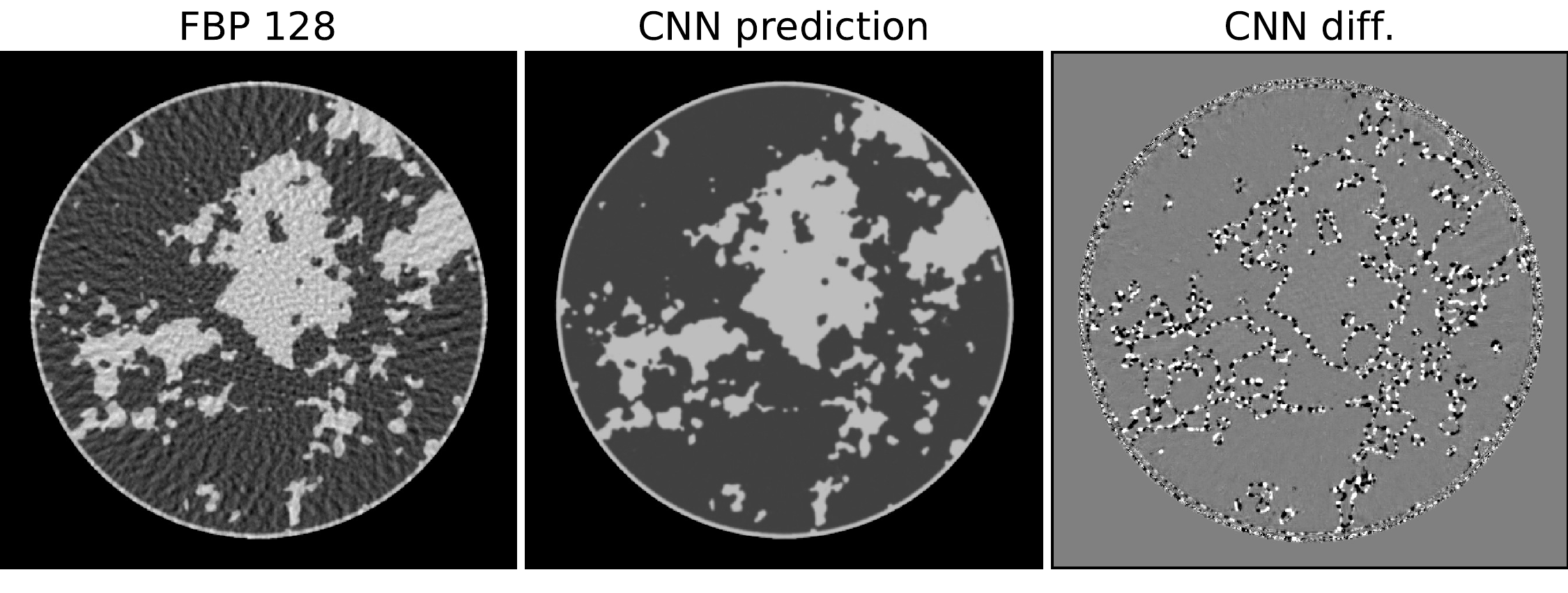}}
\centerline{\includegraphics[width=0.95\columnwidth]{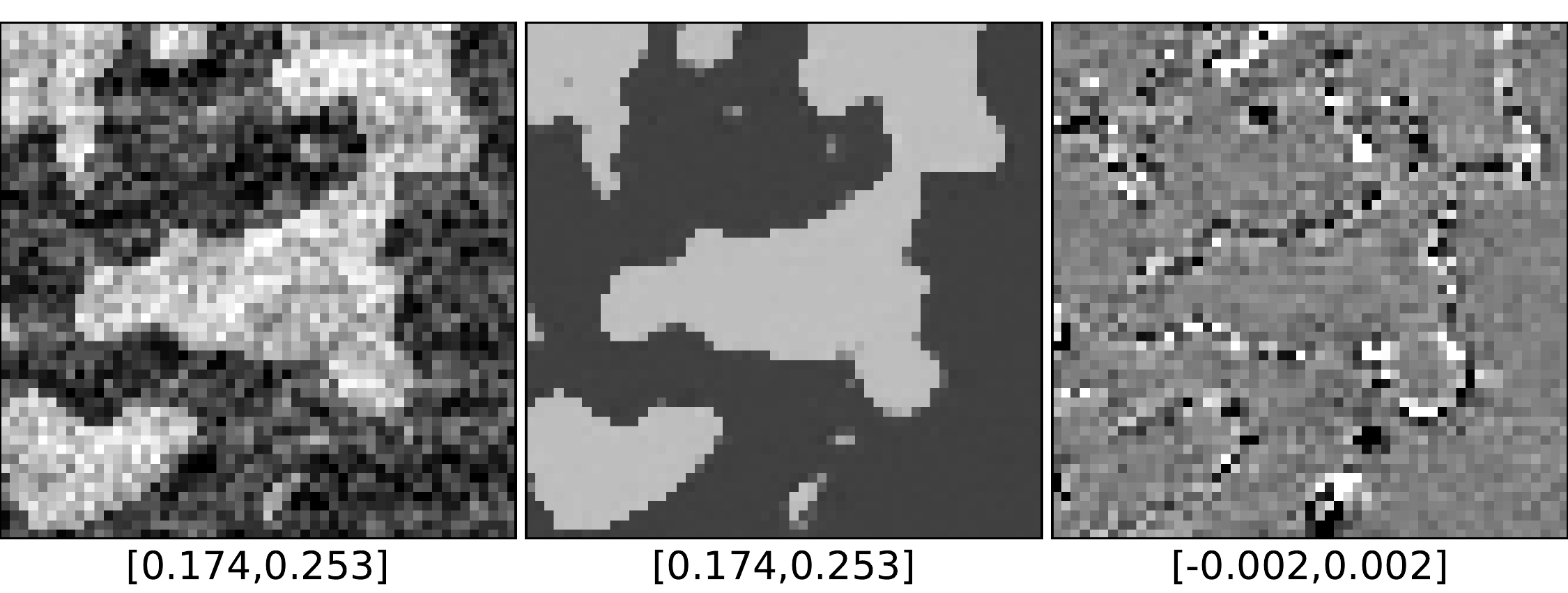}~~~
            \includegraphics[width=0.95\columnwidth]{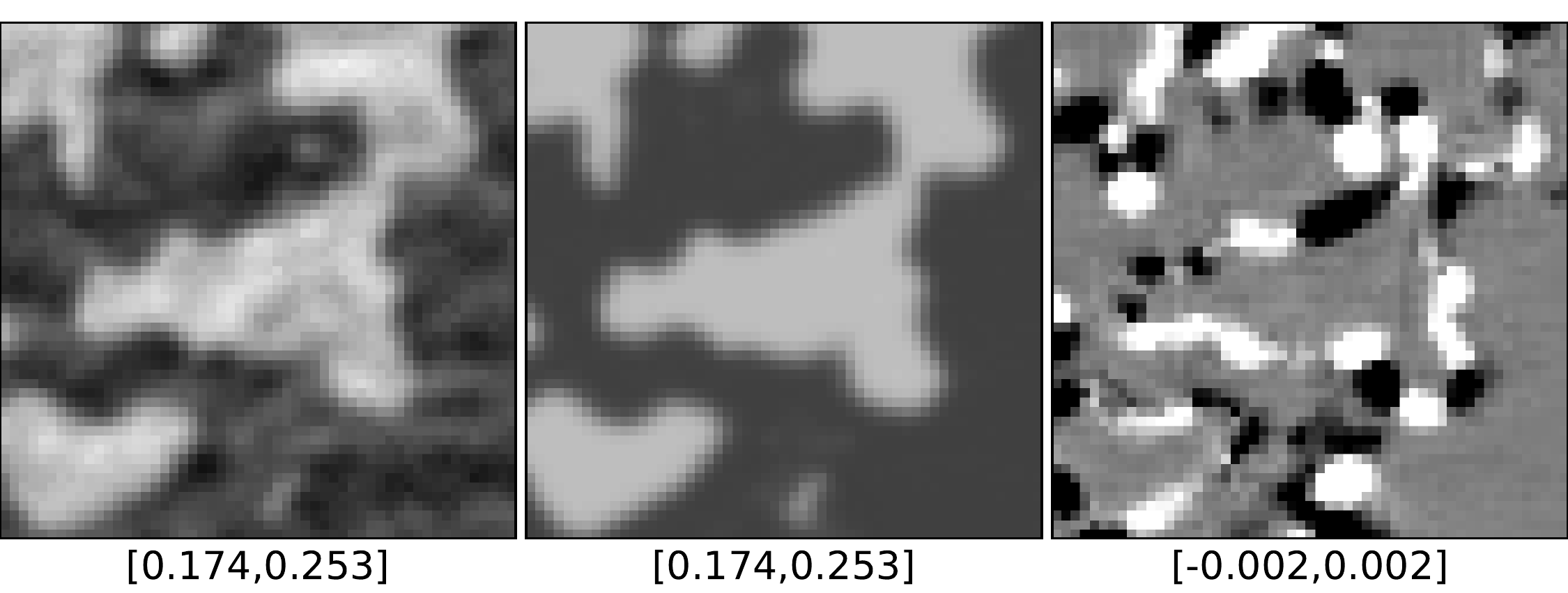}}
\caption{The left and right sets of results follow the same format as Figs. \ref{fig:binaryCNN}
and \ref{fig:smoothedgeCNN}, respectively. The difference here is that both sets of results are generated
with the same CNN, trained
with a mixture of binary and smooth-edge images and tested on an independent binary image.
The RMSE for this CNN applied to a binary test image is $1.24\times10^{-3}$ cm$^{-1}$ and
the maximum pixel error is $0.041$ cm$^{-1}$.
The RMSE for this CNN applied to a smooth-edge test image is $7.60\times10^{-4}$ cm$^{-1}$ and
the maximum pixel error is $0.0184$ cm$^{-1}$.}
\label{fig:mixedCNN}
\end{figure*}

The RMSE values for TVmin are larger than the corresponding results with
the binary phantom, but the numerical error is still orders of magnitude below the soft-tissue contrast
level of the smooth-edge phantom.
The conclusions from this study are similar to that of the binary phantom; namely, the CNN images do not
show accurate numerical recovery of the test phantom while the TVmin images do.

\subsection{Applying the CNN to training images}
\label{sec:CNNinversecrime}

As noted in Sec. \ref{sec:CNN}, the training images must belong to the restricted
set of images that are part of the learned measurement model. Testing the CNN with
data from a training image reveals the ability of the CNN to approximate the transform
from the FBP image to the reconstructed image.  We perform this test with a training
image from the binary phantom class. The results are shown on the
left set of six panels of  Fig. \ref{fig:inTS}.
Likewise, we show the result for the smooth-edge phantom class
on the right set of six panels.  In both cases, we note accurate visual recovery,
but the image reconstruction is not numerically accurate.
From an inverse problem perspective, the CNN must be able to return the true training image
from the corresponding training data in order to be able to claim that
it inverts the sparse-view CT measurement model.

\subsection{Expanding the object model}
\label{sec:expand}
In the next set of results, we investigate the impact of generalizing the object model by combining
the binary and smooth-edge phantom classes. The corresponding measurement model becomes
\begin{equation}
\label{model3}
g= R_{128} f_\text{obj} \text{ where } f_\text{obj} \in F_\text{binary} \cup F_\text{smooth-edge}.
\end{equation}
For the CNN, the training and testing set is chosen again to be 4000 image/data pairs,
where 2000 pairs are selected from the binary and smooth-edge phantoms. In this way,
the training effort is commensurate with the previous CNN results.
The sparsity exploiting reconstruction in this case requires a double optimization, TVmin over
$f$
\begin{equation}
\label{TVminw}
f^\star(w) = \argmin_f \|(|Df|_\text{mag})\|_1 \text{ such that } g = R_\text{128} G(w) f,
\end{equation}
followed by GMI sparsity optimization over $w$
\begin{equation}
\label{GMIminw}
w^\star = \argmin_w \|(|Df^\star(w)|_\text{mag})\|_0,
\end{equation}
where the $\ell_0$-norm can be efficiently optimized over $w$. Because it is a scalar argument,
the minimum can be found by bracketing and bisection.
The reconstructed image is obtained from $f^\star$  and $w^\star$ by computing
\begin{equation*}
f_\text{recon} = G(w^\star) f^\star(w^*).
\end{equation*}
Application of TVmin in this setting does require more effort, depending on how many function
evaluations are needed in the second optimization. The minimizer in the second optimization
can be accurately determined in ten to twenty iterations.

We apply the trained CNN to a binary and smooth-edge phantom in Fig. \ref{fig:mixedCNN}.
Note that in this case it is the same network
that yields both sets of results, while previously, the networks are trained with data from
the corresponding object classes. The RMSE and maximum pixel error metrics are at a level
that is similar to the previous results where all 4000 training/testing pairs came from the same class.
Thus, at least visually, the accuracy of the reconstruction seems to not be compromised, supporting
the potential for object model generalization. As the error numbers and difference images show, however,
numerically accurate inversion of the measurement model is not obtained.

\begin{figure}[h]
\centerline{\includegraphics[width=0.8\columnwidth]{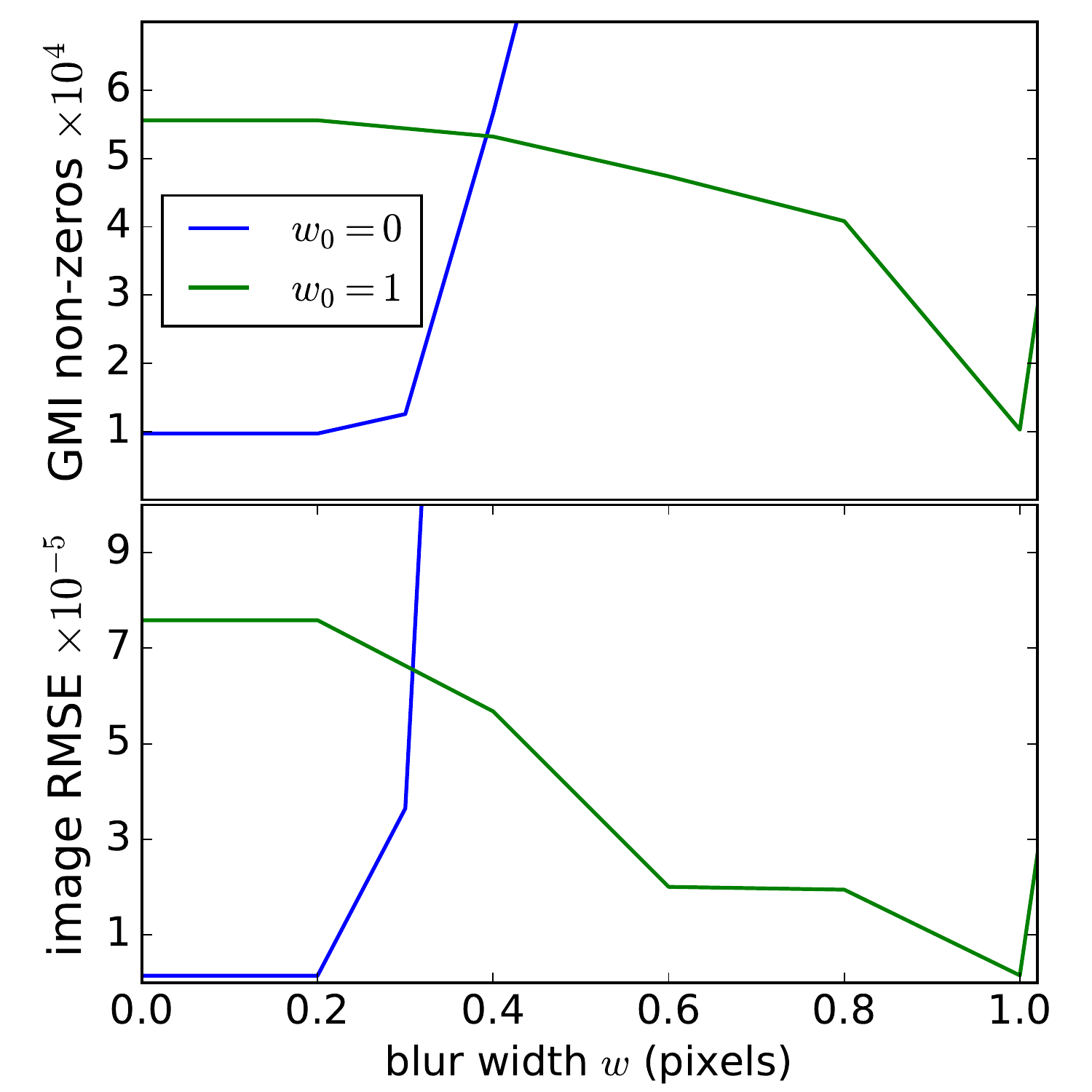}}
\caption{GMI sparsity optimization of generalized TVmin over $w$. The top panel shows the
GMI sparsity $\|(|Df^\star(w)|_\text{mag})\|_0$ as a function of $w$, where a threshold of 0.1\% of the GMI
maximum is used to distinguish non-zeros from numerically small values. The bottom row plots
the image RMSE between the smooth-edge phantom and the TVmin image estimate $G(w) f^\star(w)$
as a function of $w$.}
\label{fig:blursparsity}
\end{figure}

For the generalized TVmin results, we select a pair of phantoms from the binary and smooth-edge class; hence
the true value of $w$ are 0 and 1, respectively. The GMI sparsity optimization over $w$ is shown in
Fig. \ref{fig:blursparsity}. To perform the double optimization, the TVmin algorithm iteration
is halted when the data RMSE reaches $1 \times 10^{-6}$ so that the estimate of $f^\star(w)$
achieves similar levels of convergence as a function of $w$. The number of ``non-zeros'' in the
GMI is computed by using a threshold of 0.1\% of the maximum value in the GMI. The resulting
number of GMI pixels above this threshold is plotted in the top panel of Fig. \ref{fig:blursparsity}.
The minimum number of non-zeros occurs at the value $w=0$ and $w=1$ which coincides with the truth
values for the binary and smooth-edge images, respectively.
Having recovered the true $w$, the actual image is obtained from TVmin using $w=0$ and $w=1$ and the result
is the same as what is shown in Figs. \ref{fig:binaryTVmin} and \ref{fig:smoothedgeTVmin}, respectively.

It is also instructive to plot the image RMSE between the phantoms and the image
estimates, $G(w)f^\star(w)$, for the whole range of $w$ that is searched. The image RMSE is minimized
at the corresponding $w$-values that minimize the number of GMI non-zeros.
We do note, however, that the image RMSE is still quite low for other values of $w$;
in fact all the plotted values are less than the image RMSE obtained with the CNN. These results raise
the interesting question on how to determine whether or not other values of $w$ can still lead to
recovery of the test phantom. This issue is taken up in 
Appendix \ref{app:converge}.

\subsection{Numerical evidence for solving the inverse problem}
\label{sec:evidence}

Numerically, it is not possible to show
mathematical inversion in this setting because there is always some level of numerical error.
We have implicitly been appealing to the image RMSE or
maximum pixel error being small compared to the inherent contrast of the test phantom. That the maximum
pixel error for the TVmin result is several orders of magnitude lower than the phantom contrast implies
that there is no ROI where the difference between the true and reconstructed phantom is apparent in a gray
scale matched to the phantom contrast. Further discussion of TVmin convergence and sparsity model inversion
is presented in Appendix \ref{app:converge}.

\begin{figure}[!t]
\centerline{\includegraphics[width=0.95\columnwidth]{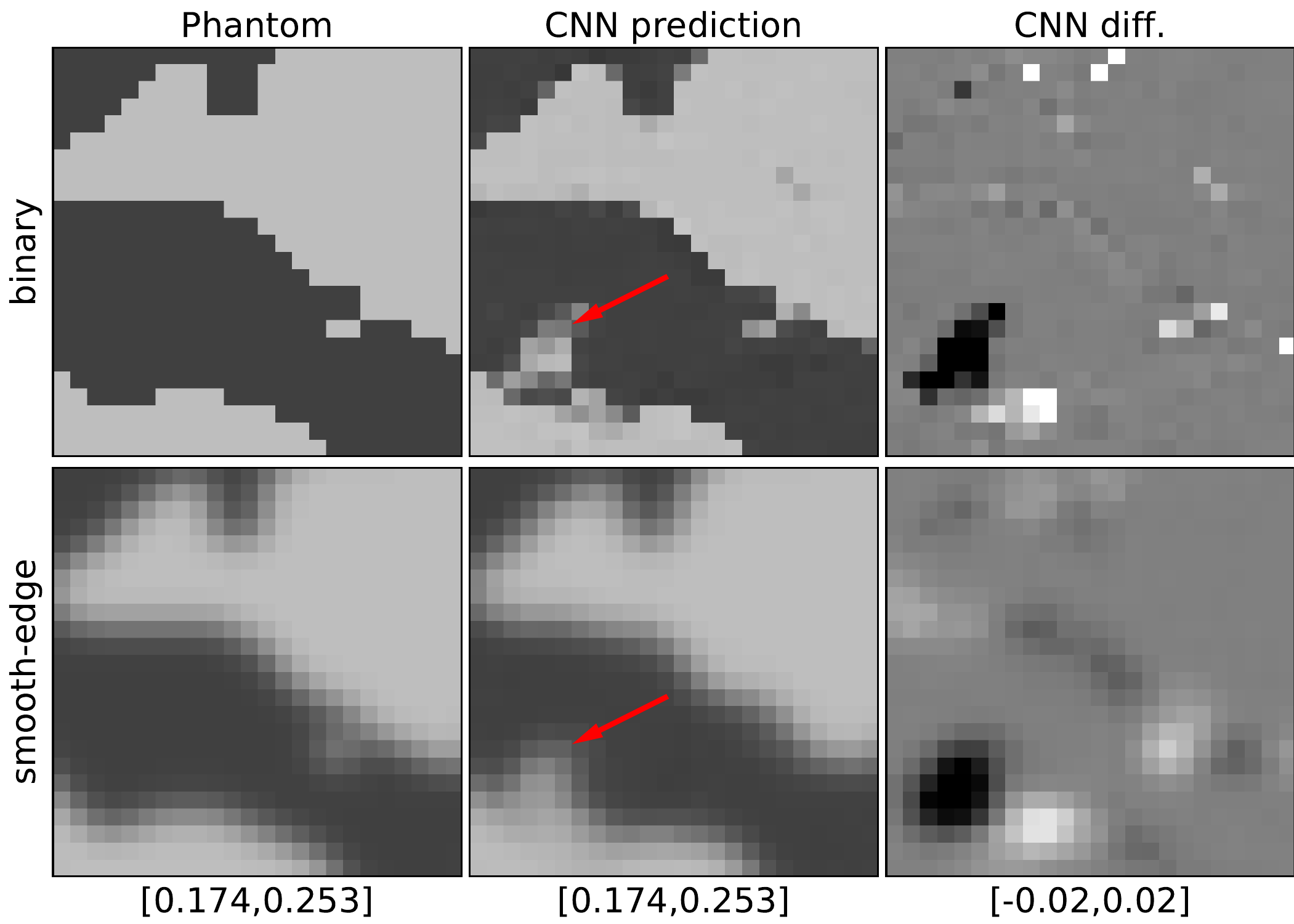}}
\caption{ROI images for the 24$\times$24 ROI with the greatest image RMSE over all ten testing
images. The top and bottom rows show the ROI images for the CNN reconstructed binary and smooth-edge
phantoms. There is visible discrepancy between the phantom and CNN reconstructed ROIs in a gray scale
appropriate for the fat/fibroglandular contrast. Most notably there are structures, indicated by
the red arrow, that are not present in the phantom ROIs. Note that the gray scale window used for the
difference image is ten times wider than the gray scale windows used for the previous CNN difference images.}
\label{fig:worst}
\end{figure}

For the CNN, we have not managed to achieve an accuracy level where the image error is imperceptible
for a gray scale matched to the phantom contrast level. To illustrate this point clearly, we have
searched all 24$\times$24 pixel ROIs in the ten testing images and selected the one with the greatest image
RMSE for both the binary and smooth-edge results presented in Secs. \ref{sec:binaryres} and \ref{sec:smoothedgeres},
respectively. The resulting ROIs are displayed in Fig. \ref{fig:worst}. On the one hand, it is a remarkable
achievement of the methodology that this is the worst ROI over ten 512$\times$512 pixel testing images.
On the other hand, there is obvious discrepancy and it is difficult to take this result as evidence for
solving a sparse-view CT inverse problem.

It may be possible that there are methods to improve the CNN accuracy.
One could imagine increasing the training set size, using
transfer learning, or an entirely different deep-learning network architecture. 
As far as we know, however,
there is no specific quantitative methodology in the literature
for estimating how much additional training, or what modifications to the deep-learning,
are needed to achieve a desired accuracy goal.

\section{Conclusion}
\label{sec:conclusion}

We have presented a focused study on solving the sparse-view CT inverse problem using
GMI sparsity-exploiting image reconstruction and deep-learning with CNNs.
The numerical evidence for solving sparse-view CT between
TVmin and the CNN is contrasted. The TVmin results have an error many orders of magnitude
smaller than that of the CNN results. In an absolute sense the scale of the TVmin
error is demonstrated to be much smaller than the inherent contrast levels of the relevant
CT application, while this is not the case for the CNN.

For TVmin there is a general inverse problem
framework for sparsity exploiting image reconstruction that underpins the algorithm
so that inverse problem results are testable and repeatable.
The CNN results show some level of accuracy, but it is not clear what it takes, nor that it is
even possible, to drive the image RMSE arbitrarily close to zero.
We have done our best to implement the residual U-net methodology described in Refs. \cite{jin2017deep,han2018framing}
but could not repeat the claim that the CT inverse problem is solved.
There remain questions:
What exactly is the measurement model that is being inverted?
How much training data is needed for a given application and system configuration?
How can the loss function be optimized so that there is arbitrarily small error on recovering the training images?
What does it take to achieve arbitrarily small error on the testing image? How exactly should the network architecture
be designed to achieve the inverse problem solution? More generally, what are all the parameters involved
in developing the CNN for inverse problems and how are they determined?

We point out that 
this negative result for the CNN methodology as an inverse problem solver employs
a much larger set of training images with perfect knowledge of the truth than would be available
from actual CT scans.
Furthermore, the training set is generated respecting gradient sparsity where we know there
exists an inverse for the 128-view CT problem as demonstrated by the TVmin algorithm.
In fact, the CNN is tested on the class of images generated by our breast model, which is
a subset of GMI-sparse images.

No other study in the literature has presented convincing evidence that CNNs solve
a CT related inverse problem.
Because this has not been demonstrated, the generalizability, conferred
by inverse problem results on injectivity and stability, cannot be applied to CNNs in the CT image
reconstruction setting.
We also point out that similar claims on using CNNs to invert full-data CT measurement models
or other inverse problems in imaging are also drawn into question by their lack of supporting evidence.

This result, however, does not mean that CNNs cannot be used for applications in CT imaging.
It only means that it is paramount to
recognize and embrace the fact that there is a high degree of variability inherent to deep-learning and CNN
methodology and, commensurately, a high degree of variability in the possible qualities of the 
CNN processed images. We must recognize and take seriously that the output of the CNN is a function
of the training set, the training methodology, network structure, and properties of the input data.
Each of these broad areas needs to be precisely specified if CNN research in the CT setting is to be repeatable.

More importantly, the images obtained from the trained CNNs must be evaluated and demonstrated
in a task-based fashion \cite{barrett2004foundations}.
Given the high degree of CNN variability and lack of mathematical characterization through
inverse problem results on injectivity and stability,
it is not clear what is the scope of applicability of a particular trained CNN. A small change in scan
configuration or object properties may or may not compromise the utility of the CNN for CT imaging tasks.
Thus, in evaluating a particular CNN, the purpose of the scan must be precisely defined and evaluated
with quantitative, objective metrics relevant to that purpose.

\appendix

\subsection{Primal-dual algorithm for solving TVmin}
\label{app:cppd}

\begin{algorithm}
\hrulefill
\begin{algorithmic}[1]
\State $f^{(k+1)} \gets f^{(k)} - \tau \left( \nu_s R^\top \lambda_s^{(k)} + \nu_g D^\top \lambda_g^{(k)} \right)$
\State $\bar{f} \gets 2f^{(k+1)} -f^{(k)}$
\State $\lambda_s^{(k+1)} \gets  \lambda_s^{(k)} + \sigma(\nu_s R \bar{f} - \nu_s g)$
\State $\lambda_g^+ \gets \lambda_g^{(k)} + \sigma \nu_g D \bar{f}$
\State $\lambda_g^{(k+1)} \gets  \lambda_g^+ / \max \left( 1, \left| \lambda_g^+ \right|_\text{mag} \right)$
\end{algorithmic}
\hrulefill
\caption{Pseudocode for the CPPD-TVmin updates. The convergence criteria are explained in the
text of Appendix A, where the additional computation of splitting variables $y_s$ and $y_g$ is needed, as
specified by Eq. (\ref{splittingvar}).
}
\label{alg:cppd}
\end{algorithm}

The Chambolle-Pock primal dual (CPPD) algorithm for solving the TVmin problem
\begin{equation}
\label{TVminapp}
f^\star = \argmin_f \|(|Df|_\text{mag})\|_1 \text{ such that } g = R f,
\end{equation}
is presented in Algorithm \ref{alg:cppd}.
The same algorithm can be straight-forwardly generalized to solve Eq. (\ref{TVminw})
where the equality constraint is replaced by $g=R G(w) f$.
To make the generalization, $R$ can be replaced by $R(w)$
\begin{equation*}
R(w) = R G(w) \text{ and } R^\top(w) = G(w) R^\top,
\end{equation*}
where we have used the fact that $G(w)$ is symmetric, i.e. $G^\top(w) = G(w)$.

The parameters and notation used in Algorithm \ref{alg:cppd} are explained here.
The parameter $\rho$ is the ratio of the primal and dual update step-size parameters, $\tau$
and $\sigma$, respectively. The integer
$k$ denotes the iteration index. The parameters $\nu_s$ and $\nu_g$ normalize the
linear transforms $R$ and $D$
\begin{equation*}
\nu_s = 1/\|R \|_2, \; \;\nu_g = 1/\|D\|_2 ,
\end{equation*}
where the $\ell_2$-norm of a matrix is its largest singular value.
The scaling is performed so that algorithm efficiency is optimized and so that results
are independent of the physical units used in implementing $R$ and $D$. Note
that the data $g$ must also be multiplied
by $\nu_s$.
The step-size
parameters $\sigma$ and $\tau$ are
\begin{equation*}
\sigma = \rho/L, \; \; \tau= 1/(\rho L),
\end{equation*}
where
\begin{equation*}
A= \binom{\nu_s R }{\nu_g D}   \; \; L= \| A \|_2,
\end{equation*}
and the matrix $A$ is constructed by stacking $\nu_s R $ on $\nu_g D$.
Due to the normalization of $R $ and $D$, and the fact that $R$ and $D$ approximately commute,
$L$ should be close to 1.
The symbol $|\cdot|_\text{mag}$ acts on the spatial vector at each pixel, yielding
the spatially dependent vector magnitude. For example, if $f$ is an image, $D f$ is the
gradient of $f$ and $|D f|_\text{mag}$ is the gradient-magnitude image (GMI).
The function ``$\max$'' acts component-wise on the vector argument.

The only free algorithm parameters are the step-size ratio
$\rho$ and the total number of iterations $K$.
The step-size ratio must be tuned for algorithm efficiency, and clearly larger $K$
leads to greater solution accuracy. For the results presented in the main text,
we set $\rho=2 \times 10^4$ and $K= 5000$ iterations.

\subsubsection*{CPPD-TVmin theory and convergence metrics}
We present a summary of the primal, saddle, and dual
optimization inherent to the CPPD framework in order to explain
the origin of the convergence criteria.
The CPPD framework is based on the use of splitting, where
Eq. (\ref{TVminapp}) is modified to
\begin{multline}
\label{splittingopt}
f^\star = \argmin_{f,y_s,y_g} \|(|y_g|_\text{mag})\|_1 \text{ such that }\\
g = y_s \text{, } Df=y_g \text{, }Rf=y_s.
\end{multline}
The introduction of the splitting variables $y_g$ and $y_s$
simplifies the potentials and constraints.
We drop the use of the normalization factors $\nu_g$
and $\nu_s$ to simplify the presentation. This equality constrained optimization is converted
to a saddle point problem by introducing Langrange multipliers for the splitting
equality-constraints 
\begin{multline}
\label{TVminsaddle1}
\max_{\lambda_s,\lambda_g} \;
\min_{f,y_s,y_g}  \|(|y_g|_\text{mag})\|_1  +\lambda_s^\top(Rf-y_s)
+\lambda_g^\top (Df-y_g) \\
\text{ such that } g=y_s,
\end{multline}
where $\lambda_s$ and $\lambda_g$ are the dual variables, a.k.a. Lagrange multipliers,
for X-ray projection and the image
gradient, respectively.
This saddle point problem can be simplified by performing the minimization over
the splitting variables $y_s$ and $y_g$ analytically
\begin{equation}
\label{TVminsaddle2}
\max_{\lambda_s,\lambda_g} \;
\min_{f}  \lambda_s^\top ( Rf -g )
+\lambda_g^\top Df
\text{ such that } \|(|\lambda_g|_\text{mag})\|_\infty \le 1,
\end{equation}
where $\|v\|_\infty=\max_i|v_i|$ yields the magnitude of the largest component of the argument vector.
It is this saddle point problem that is solved with the CPPD-TVmin algorithm.
The minimization over $f$ in Eq. (\ref{TVminsaddle2}) can also be performed analytically,
yielding the dual maximization problem to TVmin
\begin{equation}
\label{TVmindual}
\max_{\lambda_s,\lambda_g} \;
- \lambda_s^\top g 
\text{ such that } \|(|\lambda_g|_\text{mag})\|_\infty \le 1 \text{, }
D^\top \lambda_g + R^\top \lambda_s = 0.
\end{equation}

It turns out that the transversality constraint 
\begin{equation*}
D^\top \lambda_g + R^\top \lambda_s = 0
\end{equation*}
of the dual maximization in Eq. (\ref{TVmindual}) and the splitting equalities
\begin{equation}
\label{splitting}
y_s = Rf, \; \; y_g=Df,
\end{equation}
in Eq. (\ref{splittingopt})
provide a complete set of first order optimality checks.

In the CPPD framework, the splitting variable iterates can be computed from the iterates of the
primal and dual variables by adding the following two lines to the pseudocode in Algorithm \ref{alg:cppd}
\begin{align}
\label{splittingvar}
y_g^{(k+1)} &= \frac{1}{\sigma} \left(\lambda^{(k)}_g - \lambda^{(k+1)}_g\right)+D \bar{f}, \\
y_s^{(k+1)} &= \frac{1}{\sigma} \left(\lambda^{(k)}_s - \lambda^{(k+1)}_s\right)+R \bar{f}. \notag
\end{align}
A complete set of first-order convergence conditions are given by the splitting gap
\begin{equation}
\label{splittinggap}
\sqrt{\left\|y^{(k)}_s - R f^{(k)} \right\|^2_2+
\left\|y^{(k)}_g - D f^{(k)} \right\|^2_2} \rightarrow 0 ,
\end{equation}
which clearly derives from Eq. (\ref{splitting}),
and the transversality condition \cite{hiriart1993convex}
\begin{equation}
\label{transversality}
\left\|R^\top \lambda^{(k)}_s + D^\top \lambda^{(k)}_g \right\|_2 \rightarrow 0.
\end{equation}
The splitting gap is equivalent to what is called the dual residual in Ref. \cite{goldstein2013adaptive}.
These convergence criteria are particularly useful for determining the step-size
ratio parameter $\rho$, which must be tuned for algorithm efficiency.
Increasing $\rho$, increases $\sigma$ and decreases $\tau$, which has the effect
of slowing progress on decreasing the splitting gap while improving progress
toward transversality. Decreasing $\rho$ has the opposite effect.

\subsection{TVmin verification and measurement model inversion}
\label{app:converge}
The TVmin algorithm results of Sec. \ref{sec:expand} provides an opportunity to explain further the
numerical evidence for algorithm verification and measurement model inversion.
We consider the case where the test object image comes from the smooth-edge class, and we attempt
to recover the test image using Eq. (\ref{TVminw}) and
\begin{equation*}
f_\text{recon}(w) = G(w)f^\star(w),
\end{equation*}
for $w=0$ and $w=1$. The results for the image RMSE for these two $w$ values can be
read from the green curve in the bottom panel of Fig. \ref{fig:blursparsity}.
The image RMSE for $w=0$ is higher than that of the true value at $w=1$.
Yet, both image RMSEs are still small. We explore image reconstruction for these
two $w$ values further to see if the corresponding measurement models can be inverted.
The measurement model for both cases is
\begin{equation*}
g= R_{128} f_\text{obj} \text{ where } f_\text{obj} \in F_\text{smooth-edge}.
\end{equation*}
The proposed model inverse for the $w=0$ case reduces to
\begin{equation*}
f_\text{recon}(0) = \argmin_f \|(|Df|_\text{mag})\|_1 \text{ such that } g = R_\text{128} f,
\end{equation*}
and the model inverse for the $w=1$ case is the same as Eq. (\ref{TVmin128})
\begin{align*}
& f^\star = \argmin_f \|(|Df|_\text{mag})\|_1 \text{ such that } g = R_\text{128} G(w_0) f \\
& f_\text{recon}(1) = G(w_0) f^\star,
\end{align*}
where $w=w_0=1$.

\begin{figure}[!t]
\centerline{$f_\text{recon}(0)$}
\centerline{\includegraphics[width=0.75\columnwidth]{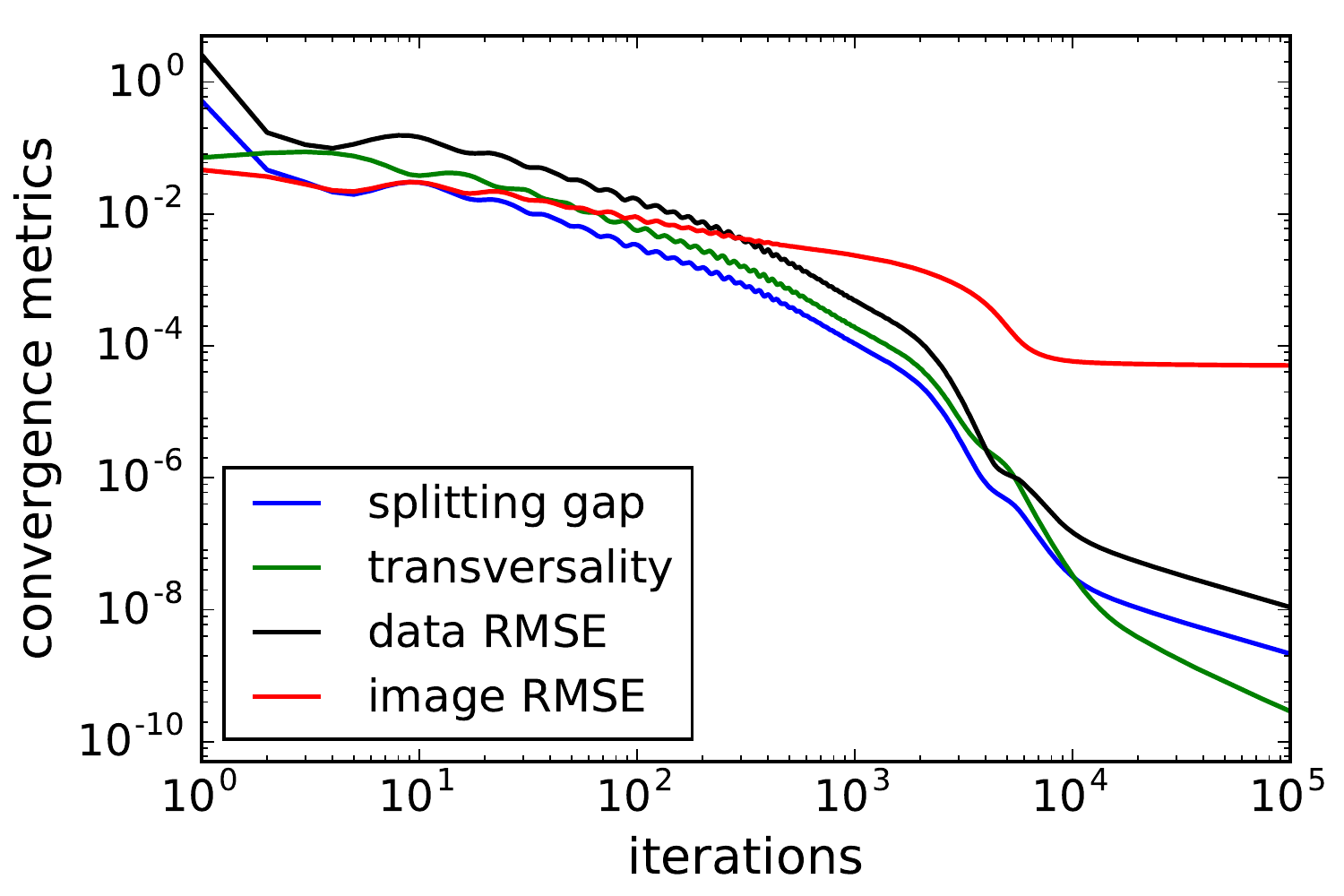}}
\centerline{$f_\text{recon}(1)$}
\centerline{\includegraphics[width=0.75\columnwidth]{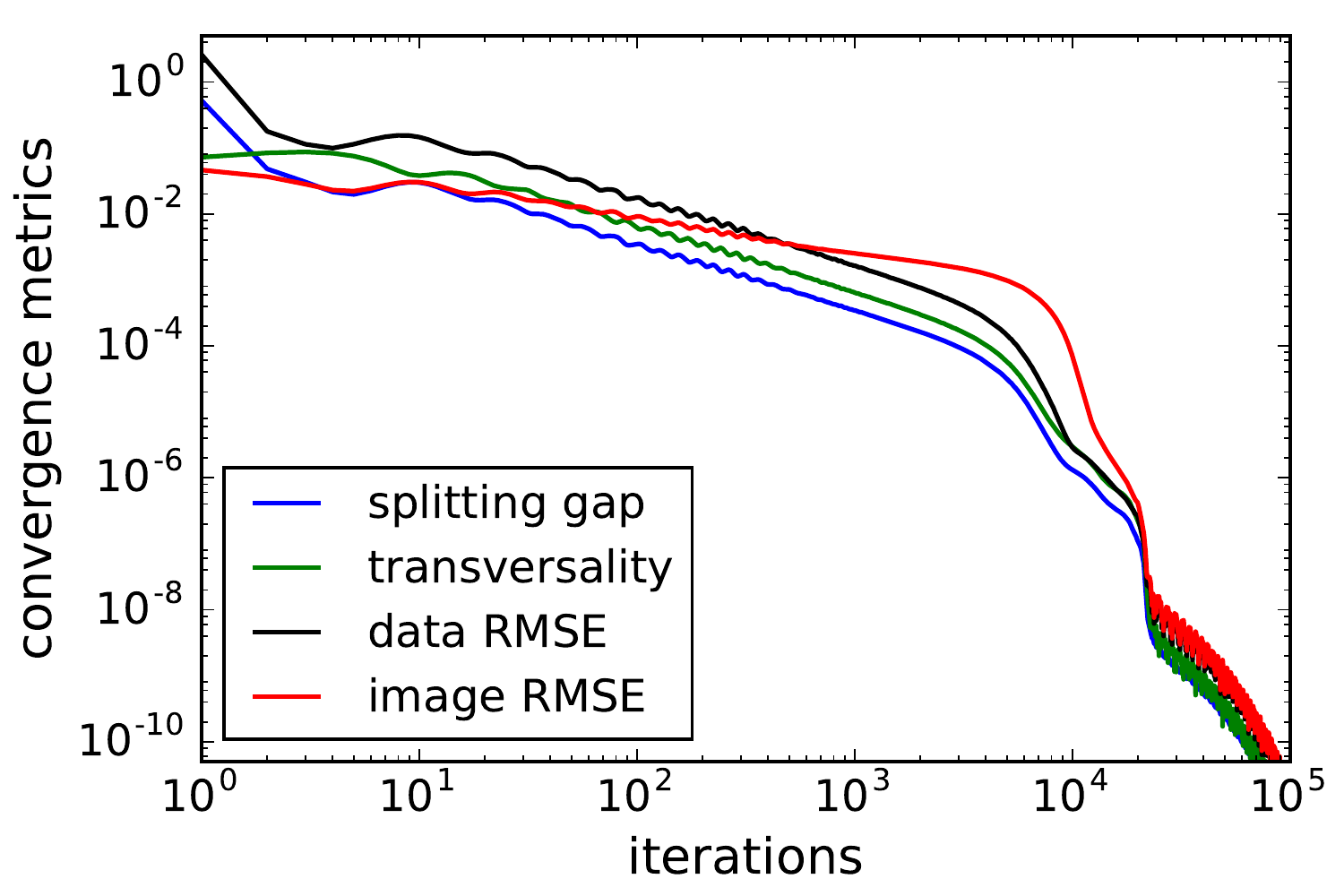}}
\caption{Plot of convergence metrics for $f_\text{recon}(w)$ when the data are generated from
projection of a smooth-edge phantom.
Shown are the image and data RMSE along with the transversality and splitting gap convergence
criteria. The latter two quantities are normalized to 1 for their value at the first iteration.}
\label{fig:convergence}
\end{figure}

From a Compressed Sensing perspective we do not expect $f_\text{recon}(0)$ to
invert the measurement model because it does not exploit the GMI sparsity optimally. The number
of non-zeros in the GMI of the smooth-edge phantom is 57050 and the number of samples in the
128$\times$512 sinogram is 65536. From empirical studies performed in Ref. \cite{jorgensen2015little}
the number of samples should be at least twice the number GMI non-zeros for this sparsity regime
and the ratio of samples to GMI non-zeros in this case is 1.15.
On the other hand, we do expect $f_\text{recon}(1)$ to invert the measurement model because
it does exploit GMI sparsity optimally as demonstrated in Sec. \ref{sec:smoothedgeres}.

For this study we extend the iteration number in computing $f_\text{recon}(w)$ to 100000 and change
the step-size ratio to $\rho=2\times10^5$.  The resulting image and data RMSE
plots are shown in Fig. \ref{fig:convergence} along with the transversality and splitting gap
convergence criteria.

The splitting gap and transversality provide the algorithm verification metrics,
and both of these curves show a downward trend verifying that the CPPD-TVmin
algorithm is solving the optimization problems associated with $f_\text{recon}(0)$ and $f_\text{recon}(1)$.
The data RMSE is a redundant algorithm verification metric in this case where
the data are ideal, but we show it also to provide reference
for the image RMSE.

It is the image RMSE that indicates whether or not the measurement model is inverted.
The image RMSE curves in Fig. \ref{fig:convergence} show a plateau for $f_\text{recon}(0)$ and a decreasing
trend for $f_\text{recon}(1)$, indicating that measurement model inversion is not attained for the former
case but is attained for the latter. Interpreting the image RMSE trend does need context, however, from
the data RMSE. Inspecting the numerical values of the image RMSE for $f_\text{recon}(0)$, they are actually
decreasing over the 100000 iterations, and considering image RMSE alone it can be difficult to distinguish
slow algorithm convergence to zero from convergence to a small positive value, i.e. plateau-ing.

When plotting the image and data RMSE together, the picture is clearer.
For the $f_\text{recon}(0)$ case there is a clear divergence between the image and data RMSE curves,
ending at a nearly four order of magnitude gap for 100000 iterations.
Accordingly, the most reasonable extrapolation for this case
is that the image RMSE limits to a small positive number and does not tend to zero.
For $f_\text{recon}(1)$, on the
other hand, the image and data RMSE track together, and the most reasonable extrapolation is that they will continue
to tend to zero as the iteration continues.
We acknowledge that these conclusions involve extrapolation of
the presented results because it is only possible to execute a finite number of iterations and
there is limitations imposed by the floating point precision.

\bibliographystyle{ieeebib}
\bibliography{cnntv}

\end{document}